\documentclass[12pt,preprint]{aastex}
\usepackage{emulateapj5}

\slugcomment{To appear in The Astrophysical Journal}

\shorttitle{Lyman Break Galaxies at $z\simeq 4$ and $5$}
\shortauthors{Ouchi et al.}

\begin{document}


\title{
Subaru Deep Survey V.\\
A Census of Lyman Break Galaxies at $z\simeq 4$ and $5$\\
in the Subaru Deep Fields: Photometric Properties\altaffilmark{1}
}


\author{Masami Ouchi        \altaffilmark{2},
        Kazuhiro Shimasaku  \altaffilmark{2,3},
        Sadanori Okamura    \altaffilmark{2,3},\\
	Hisanori Furusawa   \altaffilmark{4},
	Nobunari Kashikawa  \altaffilmark{5},
	Kazuaki Ota         \altaffilmark{5},\\
        Mamoru Doi          \altaffilmark{6},
        Masaru Hamabe       \altaffilmark{7},
        Masahiko Kimura     \altaffilmark{8},
        Yutaka Komiyama     \altaffilmark{4},\\
        Masayuki Miyazaki   \altaffilmark{2},
        Satoshi Miyazaki    \altaffilmark{4},
        Fumiaki Nakata      \altaffilmark{9},\\
        Maki Sekiguchi      \altaffilmark{10},
        Masafumi Yagi       \altaffilmark{5}, and
        Naoki Yasuda        \altaffilmark{5}
        }

\email{ouchi@astron.s.u-tokyo.ac.jp}


\altaffiltext{1}{Based on data collected at 
        Subaru Telescope, which is operated by 
        the National Astronomical Observatory of Japan.}
\altaffiltext{2}{Department of Astronomy, School of Science,
        University of Tokyo, Tokyo 113-0033, Japan}
\altaffiltext{3}{Research center for the Early Universe, School of Science,
        University of Tokyo, Tokyo 113-0033, Japan}
\altaffiltext{4}{Subaru Telescope, National Astronomical Observatory, 
        650 N.A'ohoku Place, Hilo, HI 96720, USA}
\altaffiltext{5}{National Astronomical Observatory, 
        Mitaka, Tokyo 181-8588, Japan}
\altaffiltext{6}{Institute of Astronomy, School of Science, 
        University of Tokyo, Mitaka, Tokyo 181-0015, Japan}
\altaffiltext{7}{Department of Mathematical and Physical Sciences,
        Faculty of Science, Japan Women's University, Tokyo 112-8681, Japan}
\altaffiltext{8}{Department of Astronomy, Kyoto University, 
        Sakyo-ku, Kyoto 606-8502}
\altaffiltext{9}{Department of Physics, University of Durham,
        South Road, Durham DH1 3LE, UK}
\altaffiltext{10}{Institute for Cosmic Ray Research, 
        University of Tokyo, Kashiwa, Chiba 277-8582}


\begin{abstract}

We investigate photometric properties of Lyman Break
Galaxies (LBGs) at $z=3.5-5.2$ based on large samples
of 2,600 LBGs detected in deep ($i'\lesssim 27$)
and wide-field (1,200 arcmin$^2$) images taken in 
the Subaru Deep Field (SDF) and the Subaru/XMM Deep Field (SXDF)
using broad band $B$, $V$, $R$, $i'$, and $z'$ filters.
%
The selection criteria for the LBG samples are 
examined with 85 spectroscopically identified objects, 
and the completeness and contamination
of the samples are estimated from 
Monte Carlo simulations based on a  
photometric-redshift catalog of 
the Hubble Deep Field-North.
We find that these LBG samples are nearly
rest-frame UV magnitude-limited samples, missing
systematically only 10\% of red high-$z$ galaxies (in number) 
which are a dusty population with $E(B-V)\gtrsim 0.4$.
%
We calculate luminosity functions of the LBGs
with the estimated completeness and
contamination, and find 
(i) that the number density of bright galaxies ($M_{\rm 1700}<-22$
; corresponding to SFR$\gtrsim 100 h_{70}^{-2} M_\odot$yr$^{-1}$
with extinction correction) decreases significantly from $z=4$ to $5$ and
(ii) that the faint-end slope of the luminosity
functions of LBGs may become steeper towards
higher redshifts.
%
We estimate dust extinction of $z\simeq 4$ LBGs 
with $M<M^* (\simeq -21)$ 
from UV-continuum slopes, and obtain $E(B-V)=0.15\pm 0.03$ as
the mean value.
The dust extinction remains constant with apparent luminosity, 
but increases with intrinsic (i.e., extinction-corrected)
luminosity.
%
We find no evolution in dust extinction between LBGs at $z=3$
and $4$.
%
We investigate the evolution of UV-luminosity density by
integrating the luminosity functions of LBGs,
and find that the UV-luminosity density at 1700\AA,
$\rho_{\rm UV}$ does not significantly
change from $z=3$ to $z=5$, i.e.,
$\rho_{\rm UV}(z=4)/\rho_{\rm UV}(z=3)=1.0\pm0.2$ and
$\rho_{\rm UV}(z=5)/\rho_{\rm UV}(z=3)=0.8\pm0.4$, thus
the cosmic star-formation rate (SFR)
density (with correction for dust extinction)
remains constant within the error bars,
or possibly a slight decline, from $z=3$ to $z=5$.
We estimate the stellar mass density from
the cosmic SFR thus obtained, and find that
this stellar mass density is consistent with
those derived directly from the stellar mass function
at $z=0-1$, but exceeds those at $z\sim 3$ 
by a factor of 3.
We find that the ratio of the UV-luminosity density 
of Lyman $\alpha$ emitters (LAEs) to that of LBGs is $\simeq 60$\%
at $z\simeq 5$, and thus about a half of star formation 
probably occurs in LAEs at $z\simeq 5$.
We obtain a constraint on the escape fraction of UV-ionizing photons  
(i.e., UV continuum in 900\AA) produced by LBGs, 
$f_{\rm esc} \gtrsim 0.13$. 
This implies that the escape fraction of LBGs may be larger
than that of star-forming galaxies at $z=0$.

\end{abstract}





\section{Introduction}
\label{sec:introduction}

Formation history of galaxies is basically 
understood as a combination of two fundamental evolutionary 
processes, i.e., production of stars 
and accumulation of dark matter
in the standard framework of galaxy formation
which is the Cold Dark Matter (CDM) models.
In order to investigate the formation history
of stars, one of the fundamental processes, 
many efforts have been made for searching
for high-$z$ galaxies up to $z\simeq 7$ \citep{hu2002,kodaira2003}.
Very deep optical-to-near infrared imaging data such as
the Hubble Deep Field-North (HDF-N) pioneered
to detect high-$z$ galaxies at $z\gtrsim 3$.
The photometric-redshift (photo-$z$) method accurately 
identifies the redshifts of all high-$z$ galaxies detected 
in the imaging data 
(e.g., \citealt{connolly1995,gwyn1996,lanzetta1996,sawicki1997,
wang1998,fernandez-soto1999,
benitez2000,fontana2000,furusawa2000,yahata2000,
massarotti2001,rudnick2001}).
The photo-$z$ method has an advantage in identifying
high-$z$ galaxies with little systematic selection bias.
However, the photo-$z$ method requires near-infrared images 
deep enough to detect faint high-$z$ galaxies,
as well as multi-band optical images.
Thus, application of this technique is limited to
small patches of the sky  
and to small numbers of galaxies at 
relatively low redshifts ($z\lesssim3$), 
due to the small field-of-view of near-infrared
imagers to date. 
On the other hand, the broad-band two color selection 
technique successfully identifies a large number of
high-$z$ galaxies detected only in optical bands,
which is based on Lyman-break ($912$\AA) features redshifted 
to optical wavelengths
(e.g., \citealt{steidel1996a,steidel1996b,madau1996,
lowenthal1997,madau1998,steidel1999}).
High-$z$ galaxies found by this method are
called Lyman Break Galaxies, or LBGs.
There are a few thousand of photometrically
selected high-$z$ ($2\lesssim z \lesssim 4$) galaxies,
and about 1000 galaxies have been spectroscopically
confirmed to be really located at $z\simeq 3$ \citep{steidel2003}.
The Lyman break technique is an ideal method
for selecting a large number of high-$z$ galaxies
and for studying their general properties
in a limited telescope time, since the method
requires optical images in a few bandpasses.
Furthermore, relative sensitivities in optical bands
are better than those in near-infrared, and thus
the Lyman break technique can identify higher-redshift
($z\gtrsim 4$) galaxies than the photo-$z$ method.
A problem is that
LBGs are generally UV-continuum bright galaxies,
since they are identified by their strong Lyman breaks.
It is thought that part of high-$z$ galaxies
escape from the Lyman break selection
(e.g., \citealt{pascarelle1998,dickinson2000,franx2003}). 
Care is needed when we discuss properties of
high-$z$ galaxies using LBG samples.

Luminosity function (LF) of high-$z$ galaxies are obtained to
reveal the number density and the star-formation history of galaxies
(e.g., \citealt{sawicki1997,
fernandez-soto1999,furusawa2000,
poli2001}; \citealt[hereafter SDS III]{kashikawa2003}).
\citet{steidel1999} find from their large LBG samples
that the LF of LBGs at $z=3$ is well
fitted by the Schechter function with
a steep slope, $\alpha=-1.6$, 
down to $L\simeq 0.1L^*$, and that the UV-luminosity density
derived by integrating the LF does not change from $z=3$
to $z=4$. Thus, the cosmic star-formation rate (SFR)
estimated from the UV-luminosity density
does not significantly drop at $z\gtrsim 3$, 
which is different from the early report
given by \citet{madau1996} with HDF-N galaxies.
A similar tendency in the cosmic SFR is also reported
by \citet{iwata2003} who derive the cosmic SFR at $z=5$.
Here, dust extinction is a critical issue in
estimating the cosmic SFR from the UV-luminosity density.

Dust extinction of LBGs at $z\sim 3$ has been estimated
by \citet{steidel1999,meurer1999,adelberger2000,vijh2003}
(see also \citealt{calzetti2001} for review) from UV spectral
slopes of LBGs. These authors show that the average (or median)
dust extinction of LBGs at $z\sim 3$ is $E(B-V)=0.10-0.30$,
assuming the dust attenuation law of \citet{calzetti1994}
or \citet{calzetti2000}.
Global spectral fitting from rest-frame UV to optical
SEDs of LBGs at $z\sim 3$ supports
this value \citep{sawicki1998,ouchi1999,shapley2001,papovich2001}.
Near-infrared spectroscopy has also yielded similar values of
dust extinction using the Balmer decrement
(e.g., \citealt{pettini1998,pettini2001,pettini2002}).
Since the dust-extinction correction for the UV-luminosity
ranges from a factor of 3 to 19
(presented in \citealt{vijh2003} and references therein),
it is a key to determine 
dust extinction accurately, so as
to estimate the real SFR from
the UV-luminosity density.

Measuring the stellar mass density 
\citep{brinchmann2000,cole2001,cohen2002,dickinson2003}
is an independent check for cosmic SFR measurements,
since the cosmic SFR is a derivative of 
the stellar mass density with respect to cosmic time. 
\citet{dickinson2003}
have derived the stellar mass density as a function of redshift
up to $z=3$. They have found that the stellar mass density
at $z=3$ is about 6\% of that in the present epoch.
However, the stellar mass density at $z=3$ they have estimated is less than
that derived by integrating the cosmic SFR given by \citet{steidel1999}
(see \citealt{cole2001}), 
where the cosmic SFR at high redshift ($5\lesssim z\lesssim 7$) 
is an extrapolation from measurements at $z\lesssim 4$.
It is important to obtain accurate cosmic SFRs
at $z\gtrsim 5$, and examine the cause of the discrepancy.
High-$z$ galaxies produce far-UV photons
by their star-formation activities, and
these far-UV photons contribute to
the ionization of the Universe (e.g., \citealt{madau1999}). 
Thus, it is also important to
measure the star-formation activity of high-$z$
galaxies, and investigate the relationship between
galaxies and the reionization of the Universe.

Observational studies of galaxies at
$z\gtrsim 4$ were mainly based
on data of the HDF-N and HDF-Sourth, both of which 
have only $\sim4-5$ arcmin$^2$ (about $2'\times 2'$), corresponding to 
$4\times4$ Mpc$^2$ at $z=4$ (comoving units), i.e.,
cluster scales in the present-day universe.
Surveys based on such a small area probably suffer from
cosmic variance, i.e., spatial inhomogeneities
of galaxy properties in
the Universe. 
Cosmic variance is thought to be 
one of the major ambiguities in the measurements
for luminosity function, luminosity density, and 
star-formation rate etc.
Furthermore, small-field surveys do not provide
clustering properties of galaxies on large scales
which reflect properties of dark matter in galaxies,
i.e., the other fundamental properties 
of galaxies.

In order to address the issues described above,
we carry out deep and wide-field surveys for
high-$z$ galaxies with 
Subaru Prime Focus Camera 
(Suprime-Cam; \citealt{miyazaki2002}), 
which is a wide-field ($34'\times27'$)
optical imager mounted on 8m Subaru.
We made deep and wide-field imaging for two blank fields
during the Guaranteed Time Observations (GTOs) of Suprime-Cam.
One blank field is the Subaru Deep Field
(SDF: $13^h 24^m 21.4^s$,$+27^\circ 29 ' 23''$[J2000];
\citealt[hereafter SDS I]{maihara2001};
\citealt[hereafter SDS II]{ouchi2003a}; SDS III;
\citealt[hereafter SDS IV]{shimasaku2003}), 
and the other is the Subaru/XMM Deep Survey Field
(SXDF: $2^h 18^m 00^s$,$-5^\circ 00' 00''$[J2000];
Sekiguchi et al. 2004 in preparation; see also \citealt{ouchi2001a}).
These fields have little Galactic extinction and few
bright stars over 1 deg$^2$. Thus, these fields are
suitable for studying high-$z$ galaxies by
deep observations with Suprime-Cam.
These two blank fields, SDF and SXDF, have also been observed
in two Subaru Observatory key projects.
One is the SDF Project (Kashikawa et al.). The SDF project
makes very deep observations for the SDF
with Suprime-Cam, multi-slit spectrographs 
(e.g., Subaru/FOCAS: \citealt{kashikawa2002}),
and near-infrared cameras and spectrographs
(e.g., Subaru/CISCO: \citealt{motohara2002}),
to study galaxy evolution at $z=3-7$. 
The area surveyed in the SDF project
is about 0.2 deg$^2$,
corresponding to one FoV of Suprime-Cam.
The other is the Subaru/XMM
Deep Survey (SXDS) Project (Sekiguchi et al.).
The SXDS project is a wide-field multi-wavelength survey project.
The field surveyed by Suprime-Cam in the SXDS project is about 1 deg$^2$
corresponding to 5 FoVs of Suprime-Cam.
The SXDF is observed by many instruments 
in various wavelengths;
radio with VLA, sub-millimeter with JCMT/SCUBA, infrared with
SIRTF, near-infrared with UKIRT/WFCAM, 
optical with Subaru/Suprime-Cam and Subaru/FOCAS,
ultra-violet with GALEX, and X-ray with XMM-Newton.
This paper is based on the GTO data alone, which
were taken in 2000 and 2001. Deeper images
(3-10 hours for each band) have been obtained 
in the key projects, but
reduction of them is still under way.
\footnote{
Part of the GTO data are now included
in the data of the SDF key project; e.g., about one-third of
the $i'$ and $z'$ data presented in \citet{kodaira2003}
were taken in the GTOs.
}

In this paper, we make large samples of LBGs 
at $z=4$ and $5$ with the deep and wide-field 
SDF+SXDF data 
(sections \ref{sec:observations}-\ref{sec:photometric}).
We derive LFs of LBGs from the large LBG samples,
and investigate the evolution of the LF
(section \ref{sec:luminosity}). We calculate
dust extinction for our LBGs at $z=4$, 
and examine the evolution of 
dust extinction for star-forming galaxies
over $z=0-4.5$ (section \ref{sec:dust}).
We calculate the UV-luminosity density of LBGs at $z=4$ and $5$,
and investigate the cosmic SFR, the stellar mass density, and
the escape fraction of far-UV photons for LBGs
at $z=4-5$ (section \ref{sec:evolution}) where
we use the dust extinction of LBGs obtained in the previous
section.
We present clustering properties of these LBGs 
in the companion paper \citep[hereafter SDS VI]{ouchi2003d}.
In SDS VI,
we combine the results of clustering properties
with those of photometric properties shown in this paper
using the CDM models, and give more detailed discussions.

Throughout this paper, magnitudes are in the AB system
\citep{oke1974,fukugita1995}.
The values for the cosmological parameters adopted in this paper
are: $(h, \Omega_m,\Omega_\Lambda,\Omega_bh^{2})=
(0.7,0.3,0.7,0.02)$. These values are 
the same as those obtained 
from the latest CMB observations
\citep{spergel2003}.

\section{Observations and Data Reduction}
\label{sec:observations}

\subsection{Observations}
\label{sec:observations_observations}

\subsubsection{Imaging}
\label{sec:observations_observations_imaging}

\begin{figure}
\epsscale{1.0}
\plotone{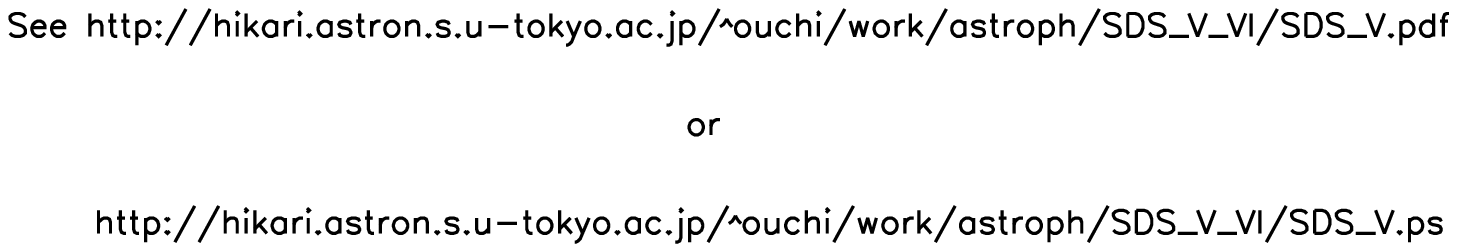}
\caption{Response of the broad-band filters
($B$,$V$,$R$,$i'$, and $z'$) which we used for
the imaging observations.
\label{fig:response_BVRiz}}
\end{figure}

During the commissioning runs 
of Suprime-Cam
(November 2000 - November 2001),
we carried out multi-band, deep and wide-field 
optical imaging in two blank fields.
One is the Subaru Deep Field
(SDF: $13^h 24^m 21.4^s$,$+27^\circ 29 ' 23''$[J2000]; SDS I),
and the other is the Subaru/XMM Deep Field
(SXDF: $2^h 18^m 00^s$,$-5^\circ 12 ' 00''$[J2000]).
The central 4 arcmin$^2$ region of the SDF
has very deep $J$ and $K'$ data (SDS I).
Although the central position of the SXDF is 
$2^h 18^m 00^s$,$-5^\circ 00 ' 00''$ (J2000) (Sekiguchi
et al. 2004 in preparation), since some bright stars are located
in the neighbor of the central position,
we choose a southern part of the SXDF which
is apart from the center by $12'$ for our observations.
We observed these two blank fields in the
$B$-, $V$-,$R$-, $i'$-, and $z'$-band filters
which cover the whole optical-wavelength range
(4000\AA\ to 10000\AA).
Figure \ref{fig:response_BVRiz} shows the response of
the filters used in these observations. The response includes
atmospheric absorption, quantum efficiency, and 
transmittance of optical elements of Suprime-Cam.

\begin{figure}
\epsscale{0.8}
\plotone{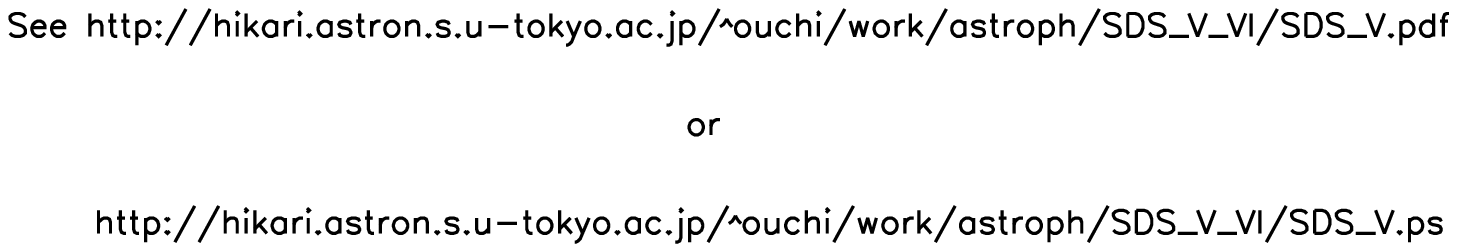}
\caption{Composite pseudo-color image of the 
Subaru Deep Field. The RGB colors are assigned to
$z'$, $i'$, and $B$ images, respectively.
The square outlined by the white line near the center
of the image is the field where deep
near-infrared $J$ and $K$ images have been obtained
by Subaru/CISCO (SDS I). A magnified
image of this field is shown 
in the upper right panel.
Gray solid line indicates the 616 arcmin$^2$ area
with good signal-to-noise (S/N) ratios in all bands except 
for the $R$ band.
We use this area for detecting $Viz$-LBGs. 
Gray dashed line corresponds to the border of 
good and bad S/N regions in the $R$ band.
The $R$ band data do not have good S/N 
below the gray dashed line, and we
use the remaining 543 arcmin$^2$ field for detecting
$BRi$-LBGs and $Riz$-LBGs. The 543 arcmin$^2$
field is also the field where we detect
87 LAEs at $z=4.86$ as described in SDS II.
\label{fig:color_sdf}}
\end{figure}

\begin{figure}
\epsscale{1.0}
\plotone{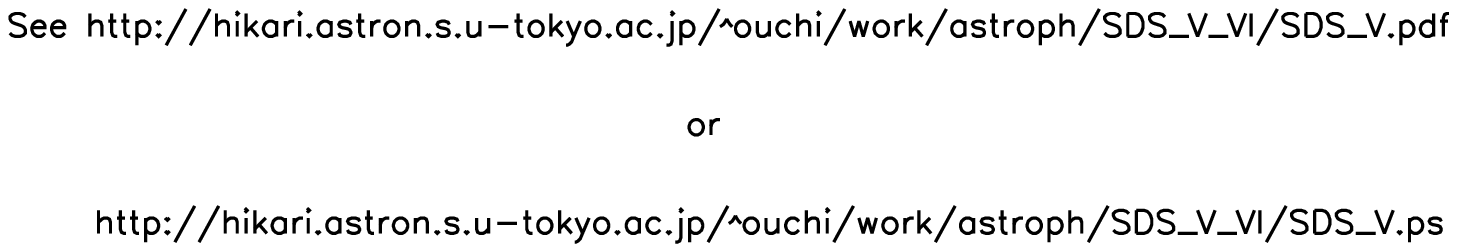}
\caption{Composite pseudo-color image of the
Subaru/XMM Deep Field. The RGB colors are assigned to
$z'$, $i'$, and $B$ images, respectively.
Gray solid line indicates the 653 arcmin$^2$ area
with good signal-to-noise (S/N) ratios.
\label{fig:color_sxdf}}
\end{figure}

The total exposure time ranges from 81 to 210 (40 to 177)
minutes among the filters in the SDF (SXDF).
The $3\sigma$ limiting magnitudes are $i'=26.9$ and $i'=26.2$
for the SDF and SXDF, respectively. During the observations,
the seeing size varied from 
$0''.5-0''.9$ and
$0''.5-1''.0$ for
the SDF and SXDF. Table \ref{tab:obs} summarizes the observations.
We show pseudo-color images of the
SDF and SXDF in
Figure \ref{fig:color_sdf} and Figure \ref{fig:color_sxdf}.
Although the FoV of Suprime-Cam is 918 arcmin$^2$ ($27'\times 34'$),
only nine CCDs were installed (i.e., one CCD named w93c2 had not 
been installed; see \citealt{miyazaki2002} for positions of
CCD chips) before March 2001.
\footnote{
In Figure \ref{fig:color_sdf}, the region in the lower right corner
surrounded by the dashed line and the solid line
corresponds to the position of the w93c2.
In Figure \ref{fig:color_sxdf}, the upper right corner, where no data
exist, corresponds to the position of the w93c2.
}
In addition, one CCD named w67c1 installed
in April 2001 showed strong fringes, and we do not use the data taken by
this CCD.
\footnote{
In Figure \ref{fig:color_sdf}, the region in the lower left corner
corresponds to the position of the w67c1.
}
We do not use, either, low-S/N regions located around the edge of 
the FoV, which are caused by dithering observations.
After we reject these bad areas, the SXDF image has 653 arcmin$^2$ 
for all bandpasses.
Similarly, the SDF image 
has 616 arcmin$^2$ for the $B$,$V$,$i'$, and $z'$ bands,
and 543 arcmin$^2$ for the $R$ band (see Table \ref{tab:obs}).

During the observations, we took images of photometric standard
stars, SA92 and SA95 \citep{landolt1992}
in the $B$, $V$, and $R$ bands,
and spectrophotometric standard stars, SA95-42 and Hz44 
\citep{oke1990} in the $i'$ and $z'$ bands.
These standard stars were observed a few times a night, when the
night was thought to be photometric.

\subsubsection{Spectroscopy}
\label{sec:observations_observations_spectroscopy}
Spectroscopic redshift data are important to check our selection
criteria for LBGs. We carried out
spectroscopic follow-up observations of LBGs
detected in our data
with Faint Object Camera and Spectrograph
(FOCAS; \citealt{kashikawa2002}) on Subaru on June 6, 2002.
We used one multi-slit mask containing slits for 
4 LBG candidates.
We also observed 5 objects selected by
other criteria including 
Lyman $\alpha$ emitter (LAE) candidates
at $z=4.9$ (SDS IV).
Thus, the number of galaxies for which we took
spectra is $4+5=9$.
We chose the 300 line mm$^{-1}$ grating with a dispersion
of 1.4 \AA\ pixel$^{-1}$ and a wavelength coverage of
4700-9400\AA. 
The sensitivity decreases in blue wavelengths ($\lesssim 6000$\AA;
see \citealt{kashikawa2002}).
We adopted a slit width of $0''.8$,
which gave a spectral resolution of 9.8\AA.
We made 2 hr exposure for each object.
The seeing size of the night was $0''.7-0''.8$.
The continuum flux limit of our spectra was 
$6.3\times10^{-19}$ erg s$^{-1}$ cm$^{-2}$ \AA$^{-1}$
with the $5 \sigma$ significance level.
In addition to these nine spectra, 
we use 76 spectra given by SDS III
which were taken during the guaranteed time observations of
FOCAS in 2001.
The details of the spectra are described 
in section \ref{sec:photometric_lbg_definitions}.

\subsection{Data Reduction}
\label{sec:observations_data}

%
%
We develop the pipeline software, SDFRED, \citep{ouchi2003c}
to reduce Suprime-Cam data.
The core programs of SDFRED are taken from
IRAF, SExtractor \citep{bertin1996}, and 
the mosaic-CCD data reduction software 
\citep{yagi2002}. SDFRED includes a 
set of optimized parameters which are
common to any Suprime-Cam data, and 
it accepts free parameters dependent on
conditions of data. This pipeline software
is open to the public, and a manual of the software 
and instructions for installation 
are given in \citet{ouchi2003c}.
We reduce all the
observed data of good quality with SDFRED, and make
stacked images for all bandpasses.
%
%
%
We align these stacked images using
hundreds of stellar objects in each image.
Then, we smooth the images with Gaussian kernels to match their seeing sizes. 
The final images have a PSF FWHM of $0''.9$
(SDF) and  $0.''98$ (SXDF).

\subsubsection{Photometric Zero-Points}
\label{sec:observations_data_photometric-zero}

%
%
%
We calculate photometric zero-points 
from photometry of standard stars with 
a $10''\phi$ aperture (Hereafter $\phi$ 
indicates the diameter of a circular aperture).
%
%
%
%
We use the zero-points obtained
in photometric nights
with airmass correction \citep{ouchi2001a}.
We check these photometric zero points using
colors of 175 Galactic stars calculated from
spectra given in
\citet{gunn1983}. Since the FoV of Suprime-Cam is large enough
to detect more than 100 bright stars with $i'\lesssim 23$, 
we can examine the photometric zero-points 
using various two-color diagrams.
We find that the colors of stellar objects in our data
are consistent with those of \citeauthor{gunn1983}'s stars,
except for some two-color diagrams in which an offset of 
$\simeq 0.05$ mag is seen.
Such offsets are probably due to 
photometric errors of
standard stars observed in slightly non-photometric conditions.
We correct the zero points by about $\simeq 0.05$ mag,
so that the observed colors of stellar objects match 
the synthetic colors of \citeauthor{gunn1983}'s stars.
The photometric-zero points thus obtained are regarded
as more accurate than 0.05 mag.

\subsubsection{Photometric Catalogs}
\label{sec:observations_data_photometriccatalogs}

Source detection and photometry are performed using
SExtractor version 2.1.6 \citep{bertin1996}.
We measure both MAG\_AUTO of SExtractor and
$2''\phi$ aperture magnitudes; 
a $2''\phi$ aperture diameter 
is twice as large as the seeing size, i.e., $1''.8$
and $2''.0$ for the SDF and SXDF data, respectively.
In order to obtain faint PSF-like objects with 
a good signal-to-noise ratio, we do not adopt
MAG\_AUTO as total magnitudes. 
Instead, we use for total magnitudes
$2''\phi$ aperture magnitudes 
after applying an aperture correction of
0.2 magnitude. We present 
in section \ref{sec:photometric_lbg_definitions}
the difference between
MAG\_AUTO and $2''\phi$ aperture magnitudes after 
aperture correction.
We make $i'$- and $z'$-detection catalogs
for both the SDF and the SXDF data.
We limit the catalogs to 
$i'<26.5$ and $z'<26.0$ ($i'<26.0$ and $z'<25.5$)
for the $i'$- and $z'$-detection catalogs of the
SDF (SXDF), in order to provide 
a reasonable level
of photometric completeness. 
These limiting magnitudes are defined for a 
$2''$ diameter aperture.
Our $i'$-detection and $z'$-detection
catalogs contain 45,923 (39,301)
and 37,486 (34,024) objects 
for the SDF (SXDF), respectively.
We correct the magnitudes of objects
for Galactic extinction, $E(B-V)=0.019$
and $E(B-V)=0.020$ \citep{schlegel1998}
for the SDF and SXDF, respectively.

We measure $3\sigma$ limiting magnitudes of the images, 
which are 
defined as the 3 $\sigma$ levels of sky noise on a $2''\phi$
diameter.
For each image, we measure sky counts in a number of 
$2''$-diameter circular apertures
which are located at randomly selected positions in the image. 
Then, we draw a histogram of the sky counts, and 
fit a Gaussian function to the histogram to obtain a $1\sigma$ noise.
When we fit a Gaussian function, we omit
the positive tail of the histogram which
is affected by objects.
The limiting magnitudes are presented in Table \ref{tab:obs}.
Note that we did not simply calculate the $3\sigma$ limiting
magnitudes by scaling a $1\sigma$ noise in one pixel to a $2''$
diameter area. Since this simple scaling results in a very optimistic
limiting magnitude as discussed in \citet{furusawa2002} and 
\citet{labbe2003},
we use the real noise values measured in $2''\phi$ apertures.

\subsubsection{Astrometry}
\label{sec:observations_data_astrometry}

Since we correct geometric distortion of images through the data reduction,
relative astrometry is good enough for this work.
Relative error is 
less than 1 pixel
($0''.202$/pixel). However, we need to obtain
absolute astrometry for follow-up spectroscopy.
We calibrate coordinates of objects using faint objects 
($B=19-20$) given in USNO A2.0 catalog \citep{monet1998}.
Although positional accuracies of USNO objects are not
so good (typically $0.\hspace{-2pt}''25$ ), there are
no other calibrators which are faint enough (so that they
are not saturated in our images).
We use 136 (97) USNO objects which are not saturated in the 
$B$-band image in the SDF (SXDF). They are uniformly distributed 
over the images. We obtain the absolute coordinates of
our objects with these USNO objects.
%
%
%
The errors in the absolute positions of objects in $(\alpha, \delta)$ 
are estimated to be less than $0.3-0.4$ arcsec
both in the SDF and in the SXDF.
%
%
%

\section{Photometric Samples of Lyman Break Galaxies at $z=3.5-5.2$}
\label{sec:photometric}

\subsection{Definitions of $BRi$-, $Viz$-, and $Riz$-Lyman Break Galaxies}
\label{sec:photometric_lbg_definitions}

We make three photometric samples of LBGs 
by the following two-color selections.
The first sample is for LBGs at $z\simeq 4$ selected by
$B-R$ vs. $R-i'$. 
They are referred to as $BRi$-LBGs.
$BRi$-LBGs are galaxies 
whose Lyman break enters
into the $B$ band and whose flat UV continuum 
is sampled in the $R$ and $i'$ bands,
and these galaxies are
identified by
their red $B-R$ and blue $R-i'$ colors.
Similarly, the second is for LBGs at $z\simeq 5$ selected by
$V-i'$ and $i'-z'$, and we refer to them as $Viz$-LBGs.
The third is for LBGs at $z\simeq 5$ selected by $R-i'$ and $i'-z'$,
and we refer to them as $Riz$-LBGs.

For $z\gtrsim4$ objects, the UV continuum 
shortward of 1216\AA\ is 
damped by the Ly$\alpha$ absorption 
of the inter-galactic medium (IGM).
Since the depression at 1216\AA\ is
as strong as that at 912\AA\ (Lyman break)
for galaxies at $z\gtrsim 4$, 
LBGs identified by the above two-color selections are
not ``Lyman break'' galaxies but ``Lyman break + 
Ly$\alpha$ absorption'' galaxies in reality.
The central redshift of these samples is shifted
toward lower redshifts than the one calculated by
dividing the central wavelength of the bluest band 
by 912\AA. For example, $Riz$-LBGs are expected to be
located around at $z\sim 6500/912-1=6.1$, but the actual
redshifts are 
$z\sim5$ (section \ref{sec:photometric_lbg_completeness}).

\begin{figure}
\epsscale{1.0}
\plotone{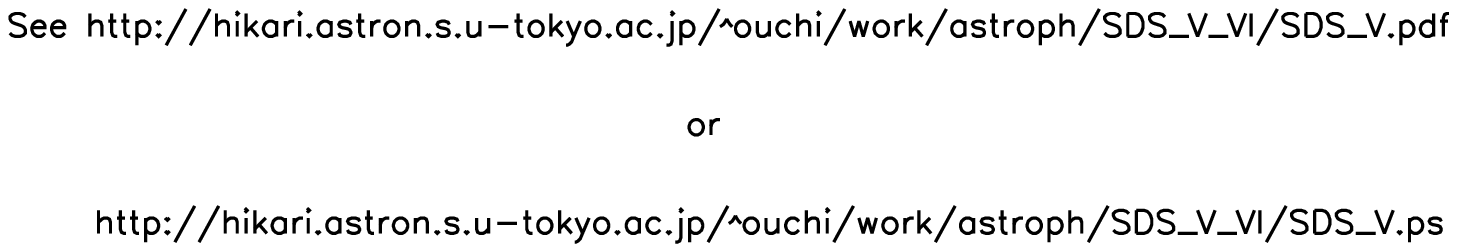}
\caption{
{\it Left panel}: $B-R$ vs. $R-i'$ color diagram displaying
the colors of model galaxies and stars. 
Red line shows the track of a
typical spectrum of star-forming galaxies from $z=3$ to $4$.
The typical spectrum is produced by
the GISSEL00 \citep{bruzual2003}
population synthesis model 
with the same parameters as
for the average $z\simeq 3$ galaxy
\citep{papovich2001}: Salpeter IMF, $Z=0.2Z_\odot$,
and 70 Myr passed from the initial
star formation 
with Calzetti's \citep{calzetti2000}
dust attenuation of $E(B-V)=0.16$. 
Filled circles on the red line indicate the
redshift from
$z=3.3$ to $z=3.9$ with an interval of $\Delta z=0.1$.
Typical spectra of elliptical, Sbc, Scd, and irregular galaxies
taken from \citet{coleman1980} are redshifted from $z=0$ to $z=3$,
which are shown by green, cyan, blue, and violet lines,
respectively.
Each line is marked by filled circles at $z=0$, $1$, and $2$.
Yellow star marks are 175 Galactic stars given by \cite{gunn1983}.
{\it Right panel}: $B-R$ vs. $R-i'$ color diagram displaying
the colors of 1048 HDF-N photo-$z$ galaxies 
obtained by convolution of their best-fit SEDs
\citep{furusawa2000}
with the filter transmissions
of Suprime-Cam. 
Black and red dots indicate 
galaxies whose photometric redshifts are 
$0<z\leq 3$ and $z>3$, respectively.
\label{fig:cc_BRi_model}}
\end{figure}

\begin{figure}
\epsscale{1.0}
\plotone{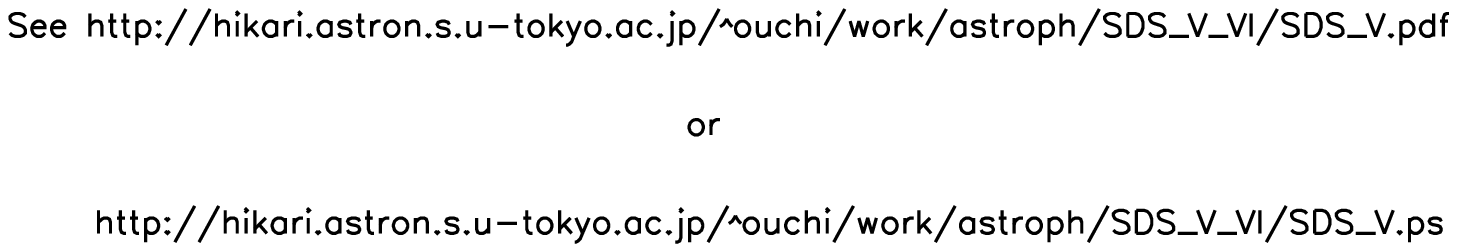}
\caption{
Same as Figure \ref{fig:cc_BRi_model}, but for
$V-i'$ vs. $i'-z'$.
Filled circles on the red line in the left panel
indicate the redshift from
$z=3.9$ to $z=4.7$ with an interval of $\Delta z=0.1$.
\label{fig:cc_Viz_model}}
\end{figure}

\begin{figure}
\epsscale{1.0}
\plotone{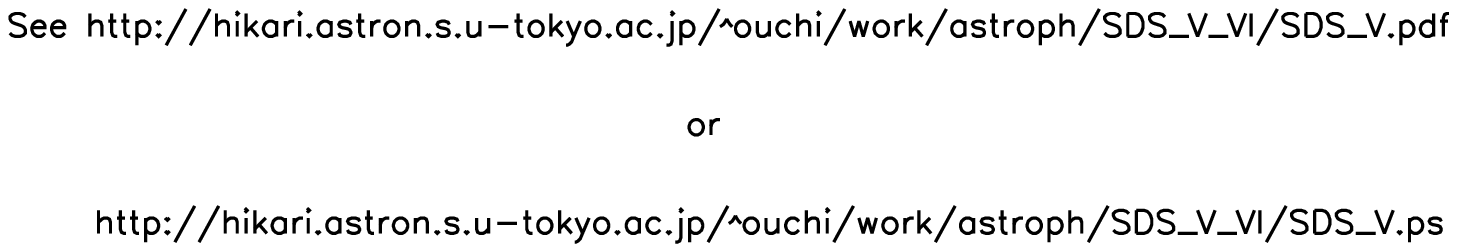}
\caption{
Same as Figure \ref{fig:cc_BRi_model}, but for
$R-i'$ vs. $i'-z'$.
Filled circles on the red line in the left panel
indicate the redshift from
$z=4.5$ to $z=5.3$ with an interval of $\Delta z=0.1$.
\label{fig:cc_Riz_model}}
\end{figure}

Figures \ref{fig:cc_BRi_model}-\ref{fig:cc_Riz_model} illustrate
how to isolate LBGs from foreground objects including Galactic
stars.
These figures indicate that LBGs can be
isolated from low-$z$ galaxies and Galactic stars
by their red break colors ($B-R$, $V-i'$, and $R-i'$)
and their blue continuum colors ($R-i'$, $i'-z'$, and 
$i'-z'$). 
In order to define the selection criteria for LBGs
quantitatively, and estimate the sample completeness
and contamination, we use 85 spectroscopically identified objects
in our fields and 1048 galaxies in the
photometric redshift catalog of HDF-N
given by \citet{furusawa2000} 
as explained in the following.

\begin{figure}
\epsscale{1.0}
\plotone{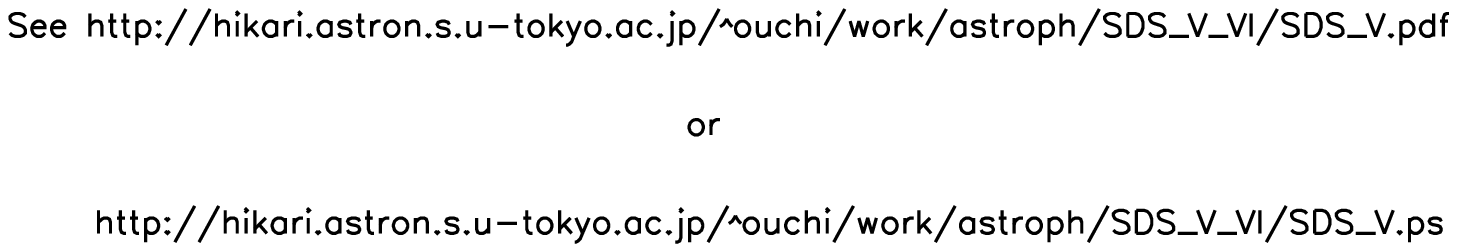}
\caption{
Spectra (left panels) and snap shots (right panels; $B$,$V$,$R$,$i'$, 
and $z'$ from left to right) 
of six Lyman break galaxies (LBGs) found in the SDF.
Object names, coordinates, and magnitudes are listed 
in Table \ref{tab:lbgs_with_redshifts}.
Bottom-left panel shows a relative night-sky spectrum.
In left panels, shaded regions correspond to 
the wavelength ranges where we cannot measure 
spectra because of bright sky emissions.
In right panels,
LBGs are located in the center of each box.
The box size is $10''\times 10''$, and
the tick marks on a side indicate $2''$.
\label{fig:spec_lbg}}
\end{figure}

Our 85 spectroscopically identified objects come from
two sources. One is the sample of 9 objects for
which we carried out spectroscopic follow-up observations
and the other is the spectroscopic catalog of 76 SDF
objects compiled by SDS III.
The nine spectra are (i) 4 LBG candidates whose colors
are close to those of model galaxies at $z>3.5$ 
(Figures \ref{fig:cc_BRi_model}-\ref{fig:cc_Riz_model}) and 
(ii) 5 objects including LAE candidates of
SDS II (see SDS IV).
We identify that these 4 LBG candidates are
real LBGs at $z=4.140$, $z=4.250$, $z=4.270$, and $z=4.865$
(These are SDFJ132413.3+274207, SDFJ132416.3+274355, 
SDFJ132413.1+274116, and SDFJ132410.5+274254 
in Figure \ref{fig:spec_lbg}). 
In the 76 spectra, there are 
two LBGs at $z>3.5$: $z=3.845$ and $4.600$.
\footnote{
The original catalog of SDS III has
four LBGs at $z>3.5$. However, one galaxy at $z=4.620$
has spectroscopic features similar to late-type stars.
Furthermore, the magnitude of this object is bright, $z'=23.66$,
and its profile is stellar. Therefore, we regard this object
as a Galactic star.
Another LBG at $z=3.810$ is blended in our image, and
we cannot detect this LBG as a single object.
Thus, we use the remaining two LBGs for our analysis.
}
%
%
%
In summary, we have 85 ($=9+76$) spectroscopically identified 
objects at $0 \le z<5$, and 6 among them are found to be 
LBGs at $z>3.5$.
Out of the rest (79 objects), three are LBGs at $3.0<z<3.5$, 
four are LAEs at $z=4.9$ which are only detected 
in a narrow-band image (SDS II), 
and the other 72 objects are either blue dwarf galaxies with emission 
lines, or red late-type galaxies up to at $z=1.1$, 
or late-type Galactic stars.
Figure \ref{fig:spec_lbg} shows the spectra and snap shots of
the identified 6 LBGs, and Table 
\ref{tab:lbgs_with_redshifts} summarizes their 
properties.
%
%
%

\begin{figure}
\epsscale{0.55}
\plotone{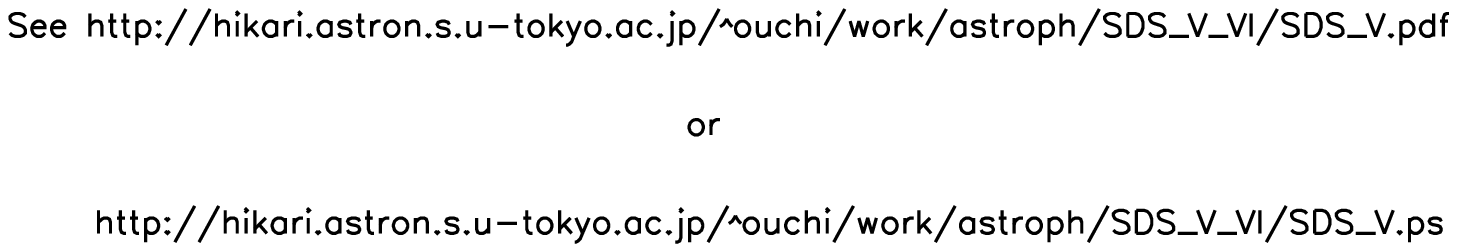}
\caption{
{\it Upper panels}: $B-R$ vs. $R-i'$ color diagram displaying
the probability maps of low-$z$ interlopers and high-$z$ galaxies
in the SDF (left panel) and SXDF (right panel).
The probability is estimated by Monte Carlo simulations 
using best-fit SEDs of galaxies in the HDF-N photometric-redshift catalog
\citep{furusawa2000}. Red and yellow contours show 
the probabilities of high-$z$ galaxies in
a bright magnitude bin ($i'=23.5-24.0$ for both fields)
and a faint magnitude bin ($i'=26.0-26.5$ for SDF, 
$i'=25.5-26.0$ for SXDF), respectively. Four contour levels are
50\%, 70\%, 90\% and 95\% (from edge to center)
of the source completeness, respectively.
The probability map of faint galaxies (yellow contours)
is generally wider than that of bright galaxies (red contours),
since fainter galaxies have larger photometric errors.
The contours for the faint galaxies do not extend to large $B-R$,
because of a limited depth of
the $B$-band data.
Blue contours show the probabilities of low-$z$ contaminants
integrated over $i'<26.5$. Their four contour levels are
0.5\%, 1.0\%, 5.0\%, and 10\% (from edge to center) of all the
contaminants in a unit area (mag$^2$; i.e.,  
$\Delta (B-R)=1$ and $\Delta (R-i')=1$).
Green line indicates the selection criteria of $BRi$-LBGs.
Red and blue circles indicate 85 spectroscopically
identified objects which are located 
at high redshift ($3.5<z<4.5$) and 
at low redshift ($0<z\leq 3.5$), respectively.
Red triangle denotes an object located at high redshift
($3.5<z<4.5$) but whose 
$B$-band magnitude is below the $1\sigma$ magnitude.
{\it Lower panels}: $B-R$ vs. $R-i'$ colors of objects
in our SDF catalog (left panel: 45,923 objects with $i'<26.5$) 
and SXDF catalog (right panel: 39,301 objects with $i'<26.0$).
A dot indicates an object. If the magnitude of an object is
fainter than the $1\sigma$ magnitude ($1\sigma$ sky fluctuation),
then the magnitude is replaced with the $1\sigma$ magnitude.
This replacement produces artificial sequences in
the two-color diagram around
$(R-i',B-R)=(1.5,1.0)$ and
$(2.0,0.7)$ for the SDF, and
$(R-i',B-R)=(1.5,1.4)$ and $(2.1,0.7)$.
for the SXDF.
Green line indicates the selection criteria of $BRi$-LBGs.
\label{fig:cc_BRi_obssim2}}
\end{figure}

\begin{figure}
\epsscale{0.55}
\plotone{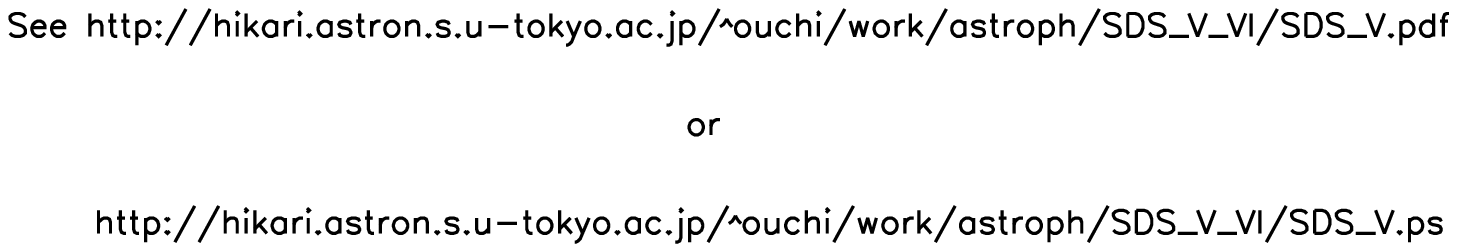}
\caption{
{\it Upper panels}: Same as the upper panels of
Figure \ref{fig:cc_BRi_obssim2},
but for $V-i'$ vs. $i'-z'$.
Red and yellow contours show 
the probabilities of high-$z$ galaxies in
a bright magnitude bin ($z'=24.0-24.5$ for both fields)
and a faint magnitude bin ($z'=25.5-26.0$ for SDF, 
$z'=25.0-25.5$ for SXDF), respectively.
Green solid lines indicate the selection criteria of $Viz$-LBGs
for the SDF data.
Green dotted line in the right panel shows the criteria of
$Viz$-LBG for the SXDF data (eq. \ref{eq:lbgselection_SXDFVizLBG}).
Red and blue circles indicate 85 spectroscopically
identified objects which are located 
at high redshift ($4.2<z<5.2$) and
at low redshift ($0<z\leq 4.2$), respectively.
Two red points far below the selection criteria
denote galaxies at $z=4.250$ and $z=4.270$.
Since the selection criteria of $Viz$-LBGs have
a completeness of just $\sim 20$\% for these redshifts,
it is consistent that the colors of these two galaxies 
are well separated from the selection criteria.
These two galaxies are selected by the criteria for $BRi$-LBGs.
{\it Lower panels}: 
Same as the lower panels of
Figure \ref{fig:cc_BRi_obssim2}, but for
$V-i'$ vs. $i'-z'$ colors of objects
in our SDF catalog (left panel: 37,486 objects with $z'<26.0$) 
and SXDF catalog (right panel: 34,024 objects with $z'<25.5$).
A dot indicates an object. If the magnitude of an object is
fainter than the $1\sigma$ magnitude ($1\sigma$ sky fluctuation),
then the magnitude is replaced with the $1\sigma$ magnitude.
This replacement produces artificial sequences in
the two-color diagram around
$(i'-z',V-i')=(1.5,1.3)$ and
$(2.2,0.4)$ for the SDF, and
$(i'-z',V-i')=(1.5,0.9)$ and $(2.0,0.3)$
for the SXDF.
Green solid lines indicate the selection criteria of $Viz$-LBGs
for the SDF data.
Green dotted line in right panel shows the criteria of
$Viz$-LBG for the SXDF data (eq. \ref{eq:lbgselection_SXDFVizLBG}).
\label{fig:cc_Viz_obssim2}}
\end{figure}

\begin{figure}
\epsscale{0.55}
\plotone{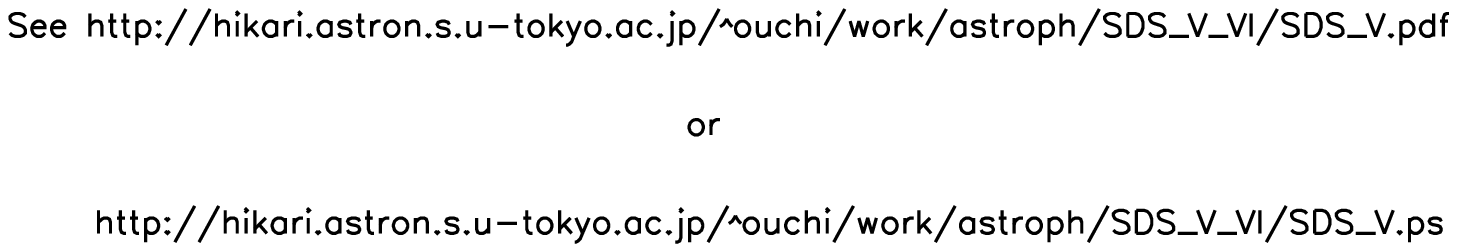}
\caption{
{\it Upper panels}: Same as the upper panels of
Figure \ref{fig:cc_BRi_obssim2},
but for $R-i'$ vs. $i'-z'$.
Red and yellow contours show 
the probabilities of high-$z$ galaxies in
a bright magnitude bin ($z'=24.0-24.5$ for both fields) 
and a faint magnitude bin ($z'=25.5-26.0$ for SDF, 
$z'=25.0-25.5$ for SXDF), respectively.
Green lines indicate the selection function of $Riz$-LBGs.
Red and blue circles indicate 85 spectroscopically
identified objects which are located 
at high redshift ($4.6<z<5.2$) and
at low redshift ($0<z\leq 4.2$), respectively.
The red point denotes a galaxy at $z=4.865$.
This galaxy is not included in the $Riz$-LBG sample
(probably due to its photometric errors),
but this galaxy is identified by the selection
criteria of $Viz$-LBGs.
{\it Lower panels}: 
Same as the lower panels of
Figure \ref{fig:cc_BRi_obssim2}, but 
$R-i'$ vs. $i'-z'$ colors of objects
in our SDF catalog (left panel: 37,486 objects with $z'<26.0$) 
and SXDF catalog (right panel: 34,024 objects with $z'<25.5$).
A dot indicates an object. If the magnitude of an object is
fainter than the $1\sigma$ magnitude ($1\sigma$ sky fluctuation),
then the magnitude is replaced with the $1\sigma$ magnitude.
This replacement produces artificial sequences in
the two-color diagram around
$(i'-z',R-i')=(1.5,1.0)$ and
$(2.2,0.2)$ for the SDF, and
$(i'-z',R-i')=(2.0,1.0)$ and $(2.0,0.0)$.
for the SXDF.
Green lines indicate the selection criteria of $Riz$-LBGs.
\label{fig:cc_Riz_obssim2}}
\end{figure}

We compare colors of galaxies in our LBG samples with 
colors of the 85 spectroscopically identified objects.
The 85 spectroscopically 
identified objects are plotted on the two-color diagrams
in Figures \ref{fig:cc_BRi_obssim2}-\ref{fig:cc_Riz_obssim2}; 
blue circles are for interlopers and
red circles are for LBGs at $3.5<z<4.5$,
$4.2<z<5.2$, and $4.6<z<5.2$ for $BRi$-LBGs,
$Viz$-LBGs, and $Riz$-LBGs, respectively.
As expected, spectroscopically identified high-$z$ galaxies
are located in the upper-left part of the
two-color diagrams.
We define the selection criteria for LBGs,
so that the criteria select LBGs
with a reasonably high completeness
and with a low contamination from interlopers.
Since the number of spectroscopically confirmed objects
is very small (especially for high-$z$ galaxies), 
we determine the selection criteria of
LBGs by simulations.

We use the best-fit SEDs of objects in the HDF-N photometric redshift catalog 
given by \citet{furusawa2000}.
The HDF-N catalog is an appropriate catalog
to be compared with our data, since it 
contains a number of galaxies at $z=4-6$ calibrated with
spectroscopic identifications.
Another advantage of the HDF-N catalog is that it has 
not only blue (in UV continuum) galaxies but also red (in UV continuum)
galaxies which usually escape from Lyman break selection
criteria.
We already show in Figures \ref{fig:cc_BRi_model}-\ref{fig:cc_Riz_model}
colors of the 1048 HDF-N galaxies
which are calculated by convolving the best-fit SEDs
with the response functions of
the Suprime-Cam filters.
Since the colors of these HDF-N galaxies are calculated 
from the best-fit SEDs,
they are free from random photometric noise.
The colors of objects in our catalogs,
on the other hand,
include random errors whose
amplitudes are dependent on 
apparent magnitudes and local sky fluctuations.
So as to evaluate these random errors,
we generate artificial galaxies that mimic the HDF-N galaxies,
and distribute them randomly on our original images
after adding Poisson noise according to their original brightness.
Then, we detect these simulated objects and measure their
brightness in the same manner as for our photometric 
catalogs (section \ref{sec:observations_data_photometriccatalogs}).
We iterate this process 100 times and derive
probability maps of the detected objects
in two-color diagrams.
We define low-$z$ interlopers
as galaxies whose redshifts are lower than $z=3$
in the original photometric-redshift catalog.
In Figures \ref{fig:cc_BRi_obssim2}-\ref{fig:cc_Riz_obssim2},
we show the probability maps of the low-$z$ interlopers, i.e.,
probability maps of contamination, thus obtained.
To derive probability maps of high-$z$ galaxies, 
we carry out additional simulations
since the number of high-$z$ galaxies in Furusawa et al.'s (2000) 
catalog is not large (52 galaxies at $z>4$).
Here, high-$z$ galaxies are defined as galaxies
whose redshifts are close to the expected central
redshift given by the color selection 
for each of the three LBG samples.
Assuming that the color distribution 
found for the high-$z$ galaxies in the HDF-N catalog is universal
and independent of $i'$ (and $z'$) magnitude, 
we make a mock catalog of 1648 galaxies at $z\geq 2.5$
whose $i'$- or $z'$-band magnitudes
are scaled from 23.0 mag to 27.0 mag 
with a 0.5-magnitude interval. Then we distribute these
galaxies on our original images and detect them in the
same manner as for the estimation of low-$z$ interlopers.
We iterate this process 100 times, and we obtain
probability maps of high-$z$ galaxies.
In Figures \ref{fig:cc_BRi_obssim2}-\ref{fig:cc_Riz_obssim2},
we show the probability maps of high-$z$ galaxies, i.e.,
probability maps of completeness.
We also plot the spectroscopically identified objects 
in Figures \ref{fig:cc_BRi_obssim2}-\ref{fig:cc_Riz_obssim2}.
The probability maps are fairly well consistent with 
the color distributions of the spectroscopically identified
objects.

Based on the probability maps of interlopers and high-$z$ galaxies, 
we determine the selection criteria of three LBG samples 
which yield small contaminants and
keep completeness high enough as:
{\footnotesize
\begin{eqnarray}
\label{eq:lbgselection_BRiLBG}
B-R>1.2,\ R-i'<0.7,\ B-R>1.6(R-i')+1.9\ \ {\rm ({\it BRi}-LBGs)},\ \ \\
\label{eq:lbgselection_VizLBG}
V-i'>1.2,\ i'-z'<0.7,\ V-i'>1.8(i'-z')+1.7\ \ {\rm ({\it Viz}-LBGs)},\ \ \\
\label{eq:lbgselection_RizLBG}
R-i'>1.2,\ i'-z'<0.7,\ R-i'>1.0(i'-z')+1.0\ \ {\rm ({\it Riz}-LBGs)},\ \
\end{eqnarray}
}
respectively.
In Figures \ref{fig:cc_BRi_obssim2}-\ref{fig:cc_Riz_obssim2},
green lines show these criteria. These criteria reject
not only the sequences of low-$z$ interlopers,
but also tails of low-$z$ interlopers scattered
by photometric errors. Since the number density of
low-$z$ interlopers is quite large, 
those scattered by photometric errors 
become significant
in number density. The upper-left panel of 
Figure \ref{fig:cc_BRi_obssim2}
shows that the selection criteria of $BRi$-LBGs
reject all the spectroscopically identified low-$z$ objects, and that
the criteria select
all the spectroscopically identified high-$z$ galaxies at $z=3.5-4.5$.
On the other hand, the upper-left panel of 
Figure \ref{fig:cc_Viz_obssim2} shows that
the selection criteria of $Viz$-LBGs reject
all the spectroscopically identified low-$z$ objects, but that
the criteria miss two spectroscopically
identified high-$z$ galaxies at $z=4.2-5.2$.
These two galaxies are SDFJ132416.3+274355 and
SDFJ132413.1+274116. Their redshifts,
$z=4.250$ and $4.270$ are 
close to the lowest redshift ($z=4.2$) 
that satisfies the criteria.
This result is consistent with our simulations
which show that the criteria select just
$\sim 20$\% of $z\simeq 4.3$ galaxies
(see Figure \ref{fig:completeness_comb}; details
are discussed later in this section).
In the upper-left panel of 
Figure \ref{fig:cc_Riz_obssim2},
the criteria of $Riz$-LBGs reject all the
spectroscopically identified low-$z$ interlopers, but that
the galaxy at $z=4.865$ is marginally missed.
Similarly, our simulations show that
the criteria select about half of 
the galaxies at $z=4.9$ (see Figure \ref{fig:completeness_comb}).
In summary, we can well estimate the completeness of 
high-$z$ galaxies as a function of redshift
from our simulations.

The right panel of Figure \ref{fig:cc_Viz_obssim2} shows that 
the $Viz$-LBG sample of the SXDF contains 
a large number of contaminants.
This is because
the signal-to-noise
ratio of the SXDF $V$-band data is worse 
than that of the SDF $V$-band data.
In order to select $Viz$-LBGs in the SXDF with a smaller number of 
contaminants, we define the $Viz$-LBG selection for
the SXDF data as
\begin{equation}
V-i'>1.2,\ i'-z'<0.7,\ V-i'>1.8(i'-z')+2.3.
\label{eq:lbgselection_SXDFVizLBG}
\end{equation}
Since these criteria are quite tight, the completeness 
of the $Viz$-LBG sample is 
very low (see Figure \ref{fig:completeness_comb}).

We apply these selection criteria to our photometric catalogs.
We use $i'$-detection catalogs for $BRi$-LBGs, 
and $z'$-detection catalogs for $Viz$-LBGs and $Riz$-LBGs.
We find 1,438 (732), 246(34), and 68 (38) objects for
$BRi$-LBGs, $Viz$-LBGs, and $Riz$-LBGs in the SDF (SXDF).
Thus we obtain LBG samples composed of 
2,556 ($\simeq 2,600$) LBGs in total, which are
the largest LBG samples at $z=4-5$, to date.
We refer to these LBG samples as the photometric samples of LBGs.
Table \ref{tab:sample} summarizes the photometric samples.

\begin{figure}
\epsscale{1.0}
\plotone{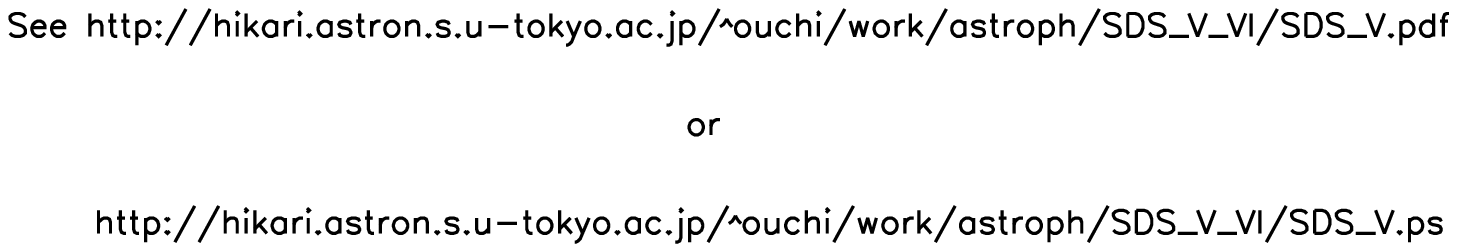}
\caption{
Number counts of the LBGs selected in the SDF and SXDF.
$BRi$-LBGs, $Viz$-LBGs, and $Riz$-LBGs are shown separately.
Magnitude, $m_{\rm AB}$, 
is $i'$-band magnitude for $BRi$-LBGs 
and $z'$-band magnitude for $Viz$-LBGs and $Riz$-LBGs.
Filled circles and filled squares are
data from the SDF and SXDF, respectively.
Since the selection criteria for $Viz$-LBGs
are different between the SDF and the SXDF,
we do not plot data for $Viz$-LBGs in the SXDF
to avoid confusion.
Open squares (top panel) and triangles (middle panel) 
indicate measurements for
LBGs at $z\simeq 4$ and $5$ obtained by 
\citet{steidel1999} and \citet{iwata2003}, respectively.
The number counts of our LBGs at $z\sim 4$ agree
well with those of Steidel et al.'s, while for $Viz$-LBGs
a large discrepancy is seen at
bright magnitudes between our measurements and 
Iwata et al.'s. See text for more details.
\label{fig:numbercount_lbg}}
\end{figure}

\begin{figure}
\epsscale{1.0}
\plotone{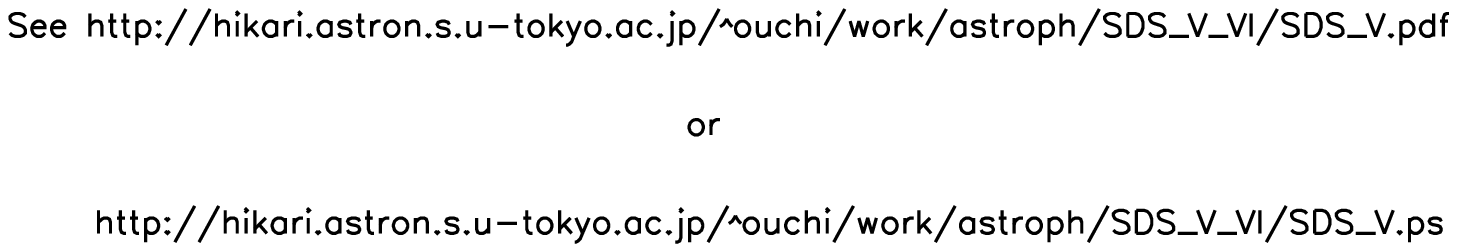}
\caption{
Completeness as a function of redshift
for our LBG samples.
In the panels of $BRi$-LBGs,
thin solid, dotted, dot-dashed,
and dashed lines denote the completeness
for $BRi$-LBGs with $i'=-\infty, 24.25, 25.25, 26.25$
($i'=-\infty, 23.75, 24.75, 25.75,$)
in the SDF (SXDF).
In the panels of $Viz$-LBGs and $Riz$-LBGs,
thin sold, dot-dashed,
and dashed lines denote the completeness
for $Viz$-LBGs and $Riz$-LBGs 
with $z'=-\infty, 24.75, 25.75$
($z'=-\infty, 24.25, 25.25$)
in the SDF (SXDF).
Thick solid lines plotted in all panels show
number-wighted completeness, which is 
calculated by averaging the
magnitude-dependent
completeness weighted by the number of
selected LBGs in each magnitude bin ($\Delta m=0.5$).
Since tighter selection criteria
(eq. \ref{eq:lbgselection_SXDFVizLBG}) are applied to 
$Viz$-LBGs in the SXDF, 
their selection window is narrower than 
that of $Viz$-LBGs in the SDF.
%
%
%
Note that the peak sensitivity of the $Riz$-LBG selection is at a
slightly lower redshift than that of the $Viz$-LBG selection.
This is because we have set a tighter selection boundary near 
the high-$z$ end ($z\simeq 5$) 
in the $Riz$-LBG selection than in the $Viz$-LBG selection 
to avoid interlopers, 
since galaxies near $z=5$ have colors
in $R-i'$ and $i'-z'$ close to low-$z$ interlopers 
(Figures \ref{fig:cc_Riz_model} and \ref{fig:cc_Riz_obssim2}). 
%
%
%
\label{fig:completeness_comb}}
\end{figure}

We show number counts of LBGs in our samples
in Figure \ref{fig:numbercount_lbg}, together with
those obtained by \citet{steidel1999} and \citet{iwata2003}.
We find that the number counts of our LBGs are fairly comparable
to those of \citet{steidel1999} who selected LBGs at $z=4.1\pm 0.5$
by a two-color diagram of $Gn-R$ vs. $R-I$. On the other hand,
there is a large discrepancy between our counts
(at bright magnitudes) and those by \citet{iwata2003}
who selected LBGs at $z=5.0\pm0.5$
by a two-color diagram of $V-I_c$ vs. $I_c-z'$. 
The reason for this discrepancy is not clear to us.
The boundary of selection criteria of \citet{iwata2003}
is very close to the colors of elliptical galaxies at $z\simeq 0.5-1.0$;
the separation between the boundary and the colors of
foreground galaxies
is only about $0.05$ magnitude in $I_c-z'$. 
In reality, the two-color diagram
of \citet{iwata2003} shows that their criteria select
a number of objects from
the outskirts of low-$z$ objects' color sequence 
in the brightest magnitude range ($I_c<23.5$; Figure 2 (d)
of \citealt{iwata2003}). 
On the other hand, 
our selection criteria
have a margin between LBGs and foreground objects 
by at least 0.2 magnitude. Photometric errors 
and small offsets in photometric zero-points 
would easily introduce errors of
0.05 magnitude.
Therefore, we infer that 
Iwata et al.'s LBG sample suffers from
bright foreground galaxies, which strongly affect the bright
end of number counts for LBGs.

\subsection{Completeness and Contamination of the Samples}
\label{sec:photometric_lbg_completeness}

\begin{figure}
\epsscale{1.0}
\plotone{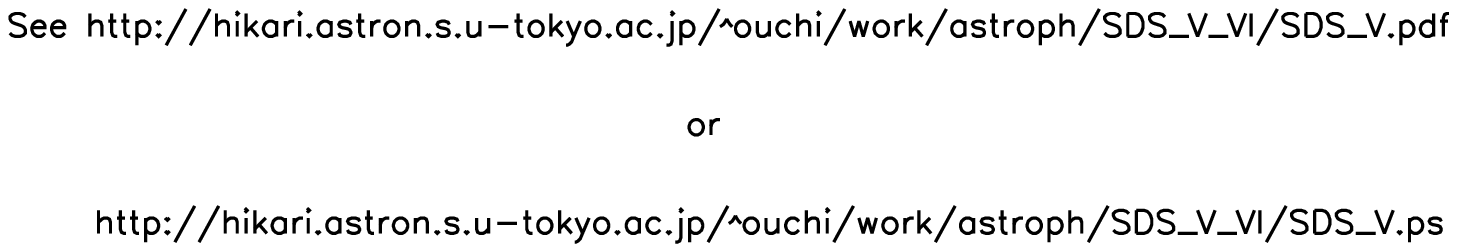}
\caption{
Number of contaminants and selected LBGs 
as a function of magnitude.
Solid lines with error-bars show the number of
all contaminants ($N_{\rm cont}^{out}$ in eq. \ref{eq:contamination}).
Dashed and dotted lines indicate the number of
low-$z$ galaxies ($N_{\rm lowz}^{out}$)
and Galactic stars ($N_{star}^{out}$).
$N_{\rm cont}^{out}$ is the sum of
these two contaminants.
The majority of the contaminants are low-$z$ galaxies for
$Viz$-LBGs in the SXDF and $Riz$-LBGs in the SDF and SXDF.
Filled circles are the numbers of objects selected
by respective LBG criteria. Magnitude, $m_{\rm AB}$, 
is $i'$ for $BRi$-LBGs 
and $z'$ for $Viz$-LBGs and $Riz$-LBGs.
\label{fig:contamination_comb}}
\end{figure}

We use the results of the simulations described
in section \ref{sec:photometric_lbg_definitions} in order to
estimate the redshift distribution, 
completeness, and contamination
of the LBG samples.
Simulated galaxies are sorted into redshift and magnitude bins
with bin sizes of $\Delta z = 0.2$ and $\Delta m = 0.5$, respectively.
We count the number of
output high-$z$ galaxies, $N_{\rm highz}^{out}(m,z)$,
and the number of input high-$z$ galaxies, $N_{\rm highz}^{in}(m,z)$,
in each of the redshift and magnitude bins.
We define the completeness as a function of redshift
for our LBG sample as
\begin{equation}
\begin{array}{rcl}
C(m,z) & = & \frac{N_{\rm highz}^{out}(m,z)}{N_{\rm highz}^{in}(m,z)}\ \ \ \ \ \ (z\geq z_0),\\
C(m,z) & = & 0\ \ \ \ \ \ \ \ \ \ \ \ \ \ \ \ \ (z<z_0),
\end{array}
\label{eq:redshiftcompleteness}
\end{equation}
where $z_0$ is the boundary redshift
between low-$z$ interlopers and LBGs. 
We adopt $z_0=3$ in our analysis.
We plot $C(m,z)$ of each LBG sample in Figure \ref{fig:completeness_comb}.
The contamination of the sample
is similarly defined as the ratio of 
the number of low-$z$ ($z<z_0$) interlopers to
the number of all the objects satisfying the selection criteria.
For contamination estimation, 
we use the total number of contaminants
lying between $z=0$ and $z=z_0$
in each magnitude bin,
because 
we do not need the number of contaminants
as a function of redshift.
Figure \ref{fig:contamination_comb} shows
the total number of contaminants
and the number of selected objects 
as a function of magnitude.
We define the contamination for a given magnitude bin as
\begin{equation}
f_{c}(m)=\frac{\int_{0}^{z_0} A N_{\rm cont}^{out}(m,z)dz}
{N_{\rm all}^{out}(m)},
\label{eq:contamination}
\end{equation}
where $N_{\rm all}^{out}(m)$ 
and $N_{\rm cont}^{out}(m)$
are, respectively, 
the total number of selected objects
and the number of selected interlopers
in the magnitude bin, and 
$A$ is a scaling factor to account for
the difference in the area between HDF-N (i.e., 
$4$ arcmin$^2$)
and the observed area (i.e., $\sim 600$ arcmin$^2$).

\subsection{Galaxies Escaping from our LBG Selections}
\label{sec:photometric_lbg_galaxies}

We examine the fraction of 
high-$z$ galaxies escaping from our LBG selections.
There are two sources by which we miss to select 
high-$z$ galaxies. One is the photometric errors,
and the other is the sample bias due to
the tight selection criteria
of LBGs which require the ideal features for
identification, i.e., a clear Lyman break
and flat-UV continuum. In the following sections,
we investigate statistical features of
our LBG sapmles.
Since the photometric errors are random errors,
correct statistical results are obtained
using the contamination and completeness given in
section \ref{sec:photometric_lbg_completeness}.
On the other hand, the sample bias
systematically changes the statistical results,
in spite of applying the corrections.
Thus it is important to know how large fractions 
of high-$z$ galaxies our LBG selections miss.

In order to investigate this sample bias,
we use the HDF-N photo-$z$ catalog by Furusawa et al. (2000) 
whose high-$z$ galaxies are not biased 
as strongly as those selected from any LBG
criteria which require a clear Lyman break
and flat-UV continuum for identification.
Since we need to know the systematic selection
bias, we use the original colors of the photo-$z$ catalog
without photometric errors.
Figure \ref{fig:photzhist_comb} plots
the redshift distribution of 
the HDF-N photometric redshift galaxies.
Galaxies which satisfy the selection criteria 
(eqs. (\ref{eq:lbgselection_BRiLBG})-(\ref{eq:lbgselection_SXDFVizLBG})) 
are shown by filled histograms. 
The second, third, and bottom panels of 
Figure \ref{fig:photzhist_comb} present the redshift distribution
of galaxies satisfying the
$BRi$-LBG, $Viz$-LBG, 
and $Riz$-LBG selection criteria, respectively.
These three panels show that most galaxies are selected
if they are located at the central redshifts of the selection windows, 
$z\simeq 4.0$ for $BRi$-LBGs,
$z\simeq 4.7$ for $Viz$-LBGs, $z\simeq 4.9$ for $Riz$-LBGs,
while the selection completeness decreases toward
the outskirts of the selection windows.
These behaviors are reasonable, since galaxies are selected as LBGs 
if their Lyman breaks and/or 
Lyman $\alpha$ breaks enter in the bluest band,
while Lyman $\alpha$ breaks do not enter
the reddest band in each of the three band sets.
The implicit assumption here is
that high-$z$ galaxies are ideally 
characterized by their
Lyman breaks and flat-UV continuum features.
However, not all galaxies share such ideal features,
and some galaxies may have weak Lyman break or
steep (red) UV continuum.
In order to estimate the fraction of
galaxies escaping from
our $BRi$-, $Viz$-, and $Riz$-LBG selections,
we show the redshifts of all galaxies
which are selected 
by at least one of eqs. (1), (2), and (3) in 
the top panel of Figure \ref{fig:photzhist_comb}.
If we focus on galaxies located at 
$z=3.9-5.1$ (47 in total) where galaxies are well selected by
the combination of $BRi$-, $Viz$-, and $Riz$-LBG selections,
we find that 4 out of the 47 escape from the
combination of these selections.
The escaping galaxies are just $4/47 \simeq 10 \%$ in number.
%
%
We investigate differences
between the four escaping galaxies and all the other galaxies
using best-fit SEDs, which reflect 
stellar population (age and metallicity) and
dust extinction ($E(B-V)$).
We find no significant difference but for
the amplitude of dust extinction.
Figure \ref{fig:missingobj_nebv} shows the histogram of 
$E(B-V)$ of galaxies in the HDF-N photo-$z$ catalog.
It is found that
galaxies with heavy dust extinction, $E(B-V)\gtrsim 0.4$,
tend to be missed. This trend is reasonable because
galaxies with heavy dust extinction are red in any colors
and these galaxies with red colors 
are outside the LBG selection criteria.

On the other hand, \citet{pascarelle1998} point out that 
about 50 \% of galaxies escape from their LBG selection. 
Their claim, however, does not conflict with ours,
since their definition of escaping galaxies is different
from ours.
Their claim is based on a comparison between $z\simeq 4$
photo-$z$ selected galaxies and galaxies selected by a LBG
selection based on
only one two-color diagram of $B-V$ vs. $V-I$.
If we calculate the completeness of our $BRi$-LBG sample
following their definition,
we find the completeness to be
61\% (36 out of 59 galaxies at $z=3.5-4.7$),
which is consistent with their value (50\%).
\citet{dickinson2000} has investigated the completeness of
LBG selection with a similar definition to ours.
He has compared
the rest-frame UV selected LBGs 
with rest-frame optical selected photo-$z$ galaxies, 
and found that 80\% of the photo-$z$ galaxies satisfy
at least one of his LBG selections. Thus his results are
consistent with ours.

%
%
%
According to the results shown above, 
we regard our LBG samples as 
{\lq}extinction-uncorrected{\rq} rest-frame UV magnitude-limited 
samples, 
missing only $\sim 10\%$ of all galaxies (in number) 
brighter than given extinction-uncorrected UV magnitudes.
Note that this very low missing fraction 
does not rule out the existence of a large population 
of very red galaxies 
whose extinction-uncorrected rest-frame UV magnitudes 
are much fainter than the limiting magnitudes of our samples 
because of heavy dust extinction.
Similarly, our result is compatible with the existence of 
passive galaxies whose UV magnitudes are fainter than 
our limiting magnitudes.
Indeed, recent deep near-infrared observations 
have revealed a substantial number of red high-$z$ galaxies 
which cannot be detected by the Lyman-break technique 
due to too faint optical magnitudes \citep{franx2003}.

\section{Luminosity Functions}
\label{sec:luminosity}

\subsection{Luminosity Functions of Lyman Break Galaxies}
\label{sec:luminosity_lbg}

 We derive luminosity functions (LFs) of LBGs at $z=4-5$ 
from our large samples;
$BRi$-LBG, $Viz$-LBG, and $Riz$-LBG samples in the SDF and SXDF.
%
%
Since we do not have redshifts for individual LBGs 
but only have the redshift distribution functions (i.e.,
probability density functions)
given in Figure \ref{fig:completeness_comb},
we cannot calculate LFs by conventional
methods (e.g., $V/V_{\rm max}$ method).
However, our LBG samples are close to volume-limited samples, 
since for each sample the width of the redshift distribution  
divided by the mean redshift of the sample is sufficiently small.
Hence, to calculate the LF for a given sample, we assign the mean 
redshift of the sample to all galaxies in it.
For each sample, using the ratio of contamination 
(eq.\ref{eq:contamination}), we calculate 
the number density, $n(m)$, of LBGs 
in a given magnitude bin by 
%
%
%
\begin{equation}
n(m)=\frac{N(m) (1-f_{c}(m))}{\int_{z_0}^{\infty} \frac{dV}{dz} C(m,z) dz},
\label{eq:numberdensity}
\end{equation}
where $N(m)$ is the number of objects
satisfying the criteria in a given magnitude bin,
$\frac{dV}{dz}$ is the differential volume, and 
$z_0(=3)$ is the boundary redshift
between low-$z$ interlopers and LBGs.
Figure \ref{fig:lumifun_comb_SFR} plots
the LFs of LBGs at $z=4.0\pm0.5$ ($BRi$-LBGs), 
$z=4.7\pm0.5$ ($Viz$-LBGs), and
$z=4.9\pm0.3$ ($Riz$-LBGs), where the abscissa is
the absolute magnitude at the rest-frame 
1700\AA. The absolute magnitude at 1700\AA\ is estimated
from the $i'$ magnitude for $BRi$-LBGs, and from the
$z'$ magnitude for $Viz$-LBGs and $Riz$-LBGs as follows. 
Most of the LBGs have PSF-like (FWHM$\simeq 1''$) shapes. 
We find from simulations made in section 
\ref{sec:photometric_lbg_definitions}
that $2''$ aperture magnitudes of PSF objects 
are fainter than the total magnitudes by 
$0.2$ mag for FWHM$\simeq 1''$ on average. 
Thus, we calculate the total magnitude 
from the $2'' \phi$ aperture magnitude, adding an 
aperture correction of $-0.2$ magnitude.
Then, we add a constant k-correction factor,
$-0.03$ for $BRi$-LBGs and $-0.01$ for $Viz$-LBGs and $Riz$-LBGs
to the total magnitude.
These values correspond to the median colors of $m_{\rm 1700}(z=4.0)-i'$,
$m_{\rm 1700}(z=4.7)-z'$, and $m_{\rm 1700}(z=4.9)-z'$, respectively,
for the HDF-N photo-$z$ galaxies
at $z=3.5-4.5$, $z=4.2-5.2$, and $z=4.6-5.2$
\citep{furusawa2000},
where $m_{\rm 1700}(z=4.0)$, $m_{\rm 1700}(z=4.7)$, and
$m_{\rm 1700}(z=4.9)$
are the magnitudes at the rest-frame
1700\AA\ for galaxies at $z=4.0$, $4.7$ and $4.9$, respectively.

We find in Figure \ref{fig:lumifun_comb_SFR} that
the LFs obtained from the SDF and the SXDF data
show an excellent consistency. Thus we regard that 
our results do not suffer from field variance.
Then we fit the Schecher function \citep{schechter1976}
{\footnotesize
\begin{equation}
\psi(M)dM=C\phi^*\exp\left\{-C(\alpha+1)(M-M^*)-\exp[-C(M-M^*)]\right\} dM,
\label{eq:schechter_m}
\end{equation}
}
to the LFs, where $C\equiv 0.4 \ln(10)$,
$\alpha$ is the power-law slope, $\phi^*$ is the normalization
factor which has a dimension of the number density of galaxies,
$L^*$ is the characteristic luminosity, and $M^*$ is the characteristic
absolute magnitude. 
%
%
%
First, we fit the Schechter function with three free parameters,
$\phi^*$, $M^*$, and $\alpha$, to the data, 
and find that these parameters cannot be constrained well 
except for the $BRi$-LBGs.
Thus, for the $Viz$-LBG LF, we fix $\alpha$ and 
determine $\phi^*$ and $M^*$ from fitting.
For the $Riz$-LBG LF, we fix $M^*$ as well as $\alpha$, 
since even two-parameter fitting ($\phi^*$ and $M^*$) 
fails to give a reliable result.
We adopt two values of $\alpha$, $-2.2$ and $-1.6$, 
for the $Viz$-LBGs and $Riz$-LBGs, 
where $\alpha=-2.2$ is the best-fit value for our $BRi$-LBGs 
and $\alpha=-1.6$ is the best-fit value for $z=3$ LBGs 
obtained by \citet{steidel1999}.
By taking a common $\alpha$ value, 
we can make a fair comparison of the luminosity density 
extrapolated beyond the detection limits 
between our samples (see section 6).
For the $M^*$ of the $Riz$-LBG LF, 
we adopt an absolute magnitude calculated 
by linearly extrapolating the $M^*$ values at $z=4.0$ ($BRi$-LBGs) 
and $z=4.7$ ($Viz$-LBGs) to $z=4.9$.
We present the best-fit parameters for the three samples in
Table \ref{tab:lumifun_schechter}, together with,
for comparison, those 
for LAEs at $z=4.86$ obtained by SDS II.
Since the Schechter fit for $Riz$-LBGs is 
quite uncertain, we refer to the results of
$BRi$-LBGs and $Viz$-LBGs 
as the Schechter parameters of $z=4$ and $z=5$ LBGs, respectively, 
in the following discussion (section \ref{sec:evolution}),
when we do not specify the sample names.
%
%
%
%
%

We plot the best-fit Schechter functions in Figure \ref{fig:lumifun_comb_SFR},
together with those of LBGs at $z=3$ and $4$ measured by 
\citet{steidel1999},
and UV-selected galaxies at $z\simeq 0$ given by 
\citet{sullivan2000}.
Figure \ref{fig:lumifun_comb_SFR} shows that the LF at $z=4$
is not much different from that at $z=3$. 
However, the number density of galaxies at faint magnitudes 
appears to be slightly higher at $z=4$ than at $z=3$, resulting in the
steeper faint-end slope, $\alpha=-2.2\pm0.2$, for our $z=4$ LBGs
than $\alpha=-1.6$ obtained by \citet{steidel1999} for $z=3$ LBGs.
Hence, the slope of the LF may be steeper at $z=4$ than at $z=3$.
Note that $\alpha$ for our LBGs is obtained from the $BRi$-LBG 
sample whose limiting magnitude reaches just $M^*+1$.
Thus the estimated slope of $\alpha=-2.2\pm0.2$ may 
include a large systematic error.

In Figure \ref{fig:lumifun_comb_SFR},
we find that the number density of bright LBGs with
$M_{\rm 1700} \sim -22$ decreases from $z=4$ to $z=5$.
This tendency is found between $BRi$-LBGs and $Riz$-LBGs
as well as $BRi$-LBGs and $Viz$-LBGs. Furthermore,
the bright $Riz$-LBGs (i.e., LBGs at $z=4.9\pm 0.3$) 
might be less numerous than the bright $Viz$-LBGs 
(i.e., LBGs at $z=4.7\pm 0.5$), implying that
the number density of bright LBGs tends to
drop at higher redshift, although the data points
of $Riz$-LBGs have a large uncertainty.
The difference between $z=4$ and $5$ is about one order
in number density at $M_{\rm 1700} \sim -22$.
Since the luminosity at 1700\AA\hspace{2pt} is
approximately proportional to the star-formation
rate (SFR), the observed decrease in
bright LBGs with $M_{\rm 1700} \sim -22$
probably indicates that galaxies with high star-formation activity
(SFR$\sim 100$ and $30 h_{70}^{-2} M_\odot$yr$^{-1}$ 
for with and without extinction correction:
details are described in section \ref{sec:discussions}) 
are rarer at $z=5$.

The error bars of the LFs in Figure \ref{fig:lumifun_comb_SFR} correspond
to Poisson errors, and we have not included errors arising from
the contamination and completeness corrections. In order to
evaluate how these corrections
affect the results, we vary the color-selection criteria
(eqs. \ref{eq:lbgselection_BRiLBG}-\ref{eq:lbgselection_SXDFVizLBG}) 
by $\pm 0.15$ magnitudes in each
color (For example,
we select $BRi$-LBGs with 
$B-R>1.2\pm0.15$, 
$R-i'<0.7\pm0.15$,
$B-R>1.6(R-i')+1.9\pm0.15)$.
We derive number counts of the selected objects and
correction factors (i.e., contamination and completeness)
with the varied color-selection criteria. Then
LFs are calculated from the number counts and
the correction factors. We find that the LFs
with the different criteria
are consistent with the LFs plotted in 
Figure \ref{fig:lumifun_comb_SFR} within a factor
$\lesssim 2$ in number density, and that the changes in criteria, 
although they give different correction factors,
do not change the three tendencies, i.e.,
no significant change from $z=3$ to $z=4$,
the decrease in number density at the bright magnitudes
from $z=4$ and $z=5$,
and the steep faint-end slope for $z=4$ LBGs.
Therefore, we conclude that these tendencies are
real.

\section{Dust Extinction}
\label{sec:dust}

\subsection{UV Slopes versus $E(B-V)$}
\label{sec:dust_uv}

Steepness of the ultraviolet spectral slope
(UV slope) of galaxies is a good measure of
dust extinction, $E(B-V)$, since
the UV continuum is sensitive to dust extinction.
\citet{meurer1999} show that 
the UV slope of local star-forming galaxies 
has a good correlation with infrared
luminosity and dust extinction obtained
from Balmer decrement (see \citealt{calzetti2001}
for a general review).
The UV slope, $\beta$, is defined as
\begin{equation}
f_\lambda \propto \lambda^\beta,
\label{eq:beta_definition}
\end{equation}
where $f_\lambda$ is the spectrum of a galaxy over the
the rest-frame wavelength of $\sim 1000$\AA\ to
$\sim 3000$\AA\, and $\beta$ is the best-fit power-low index to
the spectrum.

We estimate the dust extinction for our LBGs from the
UV slope.
We choose $i'-z'$ color for estimating UV slopes of $BRi$-LBGs 
(at $z\simeq 4$),
since $R-i'$ and $R-z'$ colors are affected
by Lyman $\alpha$ emission and 
the trough of Lyman $\alpha$ absorption systems 
while $i'-z'$ samples the continuum at $>1216$\AA.
We do not calculate UV slopes for $Viz$- or $Riz$-LBGs
since Lyman $\alpha$ emission and 
the trough of Lyman $\alpha$ absorption systems 
enter the $i'$ band.
%
The UV slope, $\beta$, is originally defined as
the index of the best-fit power-law (eq. \ref{eq:beta_definition})
in the wavelength range of $\sim1000$\AA\ to $\sim3000$\AA,
The definition of the wavelength range, however,
varies among authors, e.g., $1100-3000$\AA\ 
and $1200-1800$\AA\ etc. (see \citealt{calzetti2001}).
In order to measure the UV slope of LBGs at $z=4$ from
the $i'-z'$ color, 
we define $\beta_{iz}$ as

\begin{eqnarray}
\beta_{iz} & \equiv & - \frac{i'-z'}{2.5\log (\frac{\lambda_c(i')}{\lambda_c(z')})} \nonumber \\
           & =      & 5.42(i'-z'), 
\label{eq:beta_iz}
\end{eqnarray}
where $\lambda_c(i')$ and $\lambda_c(z')$ are the central wavelengths
of the $i'$ and $z'$ bands.

We use the latest version of \citet{bruzual1993} stellar 
population synthesis model (GISSEL00; \citealt{bruzual2003}) 
convolved with the
extinction curve of \cite{calzetti2000}. 
\footnote{
Note that the Calzetti's attenuation curve
provides the relation of
$A(1600)=k_{\rm 1600} E(B-V)$, and 
$A(1700)=k_{\rm 1700} E(B-V)$, where
$k_{\rm 1600}=10$ and $k_{\rm 1700}=9.6$.
}
We generate 
a template spectrum whose model parameters 
have the average values observed for
LBGs at $z=3$,
which are given in \citet{papovich2001}, that is, 70 Myr age,
$0.2 Z_\odot$ metallicity, and Salpeter IMF.
Figure \ref{fig:comp_meurer1999} compares 
the extinction-$\beta$ relation of
the template spectrum with 
the measurements for 43 nearby
galaxies \citep{meurer1999}.
The fitting is made over $1250-2600$\AA\ for both
the template spectrum and for the nearby galaxies.

The extinction-$\beta$ relation from
our template spectrum (solid line) shows fairly good agreement
with the data points of the nearby galaxies (filled circles), and
with the best-fit line (dashed line) derived by \citet{meurer1999}.
The template, the typical LBG spectrum convolved
with the dust extinction curve, is found to reproduce
the empirical relation of nearby starburst galaxies.
For the readers' eye guide,
two different models
are also plotted in Figure \ref{fig:comp_meurer1999}; 
one has the same age
as the template model ($70$Myr) but a higher metallicity
of $1 Z_\odot$, and the other has the same metallicity
as the template model
($0.2 Z_\odot$) but a younger age of $90$ Myr.

Then, we calculate the $E(B-V)-\beta_{iz}$ relation
using the template model. 
The solid line of 
Figure \ref{fig:redshift_beta} shows
the relation for LBGs at $z=4$, which
is expressed by a linear function:
\begin{equation}
E(B-V)=a+b \beta_{iz},
\label{eq:ebv_beta}
\end{equation}
where $a=0.0162$ and $b=0.218$. 
Here $i'$ and $z'$-band magnitudes correspond to average fluxes
in the rest frame of $1500\pm150$\AA\ and 
$1800\pm130$\AA\ for LBGs at $z=4$.
Note that
$i'$ and $z'$-band magnitudes measure the rest-frame fluxes of
$1700\pm170$\AA\ and $2000\pm140$\AA\ for LBGs at $z=3.5$, and 
$1400\pm140$\AA\ and $1600\pm110$\AA\ for LBGs at $z=4.5$.
Since $\beta_{iz}$ is 
defined in the observed frame,
this relation depends
on the redshift of galaxies.
We calculate $\beta_{iz}$ for
the template model at $z=3.5,4.5, 4.7$, and $5.2$,
where we apply the absorption of 
the Lyman $\alpha$ forest \citep{madau1995}
for the continuum flux at $<1216$\AA.
Figure \ref{fig:redshift_beta} 
displays the redshift dependence, showing
that the dependence is only $\pm 0.05$ in $E(B-V)$
for galaxies with $E(B-V) \lesssim 0.5$ at
$z=3.5-4.5$, 
but that it exceeds 
$\Delta E(B-V) \simeq 0.1$ for galaxies at $z\gtrsim 4.7$.
This large dependence for high-$z$ galaxies
is caused by Lyman $\alpha$ absorptions
in the IGM; 
the Lyman $\alpha$ forest starts entering in the blue wavelength of
the $i'$-band response function for objects at $z\gtrsim 4.7$, resulting
in a systematic reddening in $i'-z'$ color (hence, 
a systematic increase in $\beta_{iz}$).
At $z=3.5-4.5$,
the $\beta_{iz}$ estimation is 
not affected either by the Lyman $\alpha$ emission line at $1216$\AA,
or by the continuum emission from an
older stellar population 
(e.g., F stars) at $\gtrsim 2700$\AA.
Therefore, eq. (\ref{eq:ebv_beta}) holds
for LBGs at $z=3.5-4.5$ with an accuracy of 0.05 mag in $E(B-V)$,
but the equation does not work for LBGs at $z\gtrsim 4.7$. 
This is the reason why
we estimate $E(B-V)$ for $BRi$-LBGs at $z=3.5-4.5$ alone.

\subsection{Dust Extinction of Lyman Break Galaxies at $z=4$}
\label{sec:dust_dust}

We calculate the extinction of our LBGs with
eqs. (\ref{eq:beta_iz}) and (\ref{eq:ebv_beta}).
\footnote{
Combination of equations (\ref{eq:beta_iz}) and (\ref{eq:ebv_beta})
gives a relation between $E(B-V)$ and $i'-z'$ of
$E(B-V)\simeq 0.0162+1.18(i'-z')$.
}
We show the histogram of estimated $E(B-V)$
in Figure \ref{fig:beta_BRi_dist_comb}.
Because our Lyman-break selections identify galaxies with $E(B-V)\lesssim0.5$
as shown in Figure \ref{fig:missingobj_nebv}, we think that 
the estimated $E(B-V)$ values larger than 0.5 are spurious.
In Figure \ref{fig:beta_BRi_dist_comb}, we limit our sample to 
651 LBGs brighter than
$i'=25.5$ for the SDF and $i'=25.0$ for the SXDF.
This is because we need to measure colors 
up to $i'-z'\simeq 0.5$, which corresponds
to $E(B-V)\simeq 0.5$, and the sample galaxies should have
$i'$ magnitudes brighter by 0.5 mag
than the limiting magnitude of the $z'$-band image
($z'_{\rm lim}=26.0$ for the SDF and $z'_{\rm lim}=25.5$ for the SXDF).
However, there are still systematic biases in $i'-z'$ in our LBG
sample.
One is that the LBG criteria select
bluer galaxies more preferably at fainter magnitudes, since the LBG selection
needs a red $B-R$ color for the identification. This effect
is also seen in the results of simulations as the difference in the
red and yellow contours of the top panels 
in Figure \ref{fig:cc_BRi_obssim2}.
The other effect is that the edge of the $BRi$-LBG criteria 
at $1.2<B-R<3.0$ tends to reject red LBGs at 
redshift lower than $z\simeq4$. So as to correct these biases, 
we carry out Monte Carlo simulations similar to
those described in section \ref{sec:photometric_lbg_completeness}.
We use the template model of LBGs at $z=4$ 
to calculate $B-R$ and $R-i'$ colors by varying extinction
over $0.0<E(B-V)<1.0$. We then make artificial 
galaxies which mimic the colors of the templates, 
and distribute them in the original
images. We detect the artificial galaxies, and select them 
with the $BRi$-LBG criteria, and derive completeness as a function of 
$E(B-V)$ for four $i'$ magnitude bins.
The results are plotted in the top
panel of Figure \ref{fig:beta_BRi_dist_comb}. 
Using these completeness functions, we derive completeness-corrected
distributions of $E(B-V)$, which are shown by shaded histograms 
in Figure \ref{fig:beta_BRi_dist_comb}.

The mean values of $E(B-V)$ calculated
from the completeness-corrected histograms are
0.18, 0.13, 0.15, and 0.14
for magnitude bins of
$i'=23.5-24.0$, $24.0-24.5$, $24.5-25.0$, and
$25.0-25.5$, respectively (Figure \ref{fig:beta_BRi_dist_comb}).
The mean extinction
of all the $BRi$-LBGs with $i'<25.5$
is $E(B-V)=0.15\pm0.03$. Thus the average extinction of
LBGs at $z\simeq 4$ is estimated with the Calzetti's law
to be about a factor of 4 for luminosity in 1700\AA.
Since $i'=25.5$ corresponds to $M_{\rm 1700}=-20.8$
whose magnitude is comparable to $M^*_{\rm 1700}=-21.0-$ $-20.6$
(Table \ref{tab:lumifun_schechter}), the mean extinction
of $E(B-V)=0.15\pm0.03$ is for LBGs with $M\lesssim M^*$.
No significant dependence of $E(B-V)$ on the magnitude is
found over $23.5<i'<25.5$ (corresponding to $-23<M_{\rm 1700}<-21$).
Note that these magnitudes are apparent 
(i.e., before extinction correction) magnitudes, and this
result indicates that LBGs at $z=4$ have 
no significant correlation between $E(B-V)$ and the apparent magnitude.
\citet{adelberger2000} have obtained a similar result for LBGs at $z=3$.
It is, however, found that 
there is a correlation between dust extinction and 
(extinction-corrected) intrinsic luminosity for our LBGs.
Figure \ref{fig:beta_BRi_dist_comb} shows that for each magnitude bin, 
$E(B-V)$ spans the range of $0\lesssim E(B-V) \lesssim 0.5$, 
which is larger than the statistical errors. 
This means that intrinsically brighter LBGs are generally
more heavily attenuated by dust. The same tendency is found
in LBGs at $z=3$ by \citet{meurer1999,shapley2001}.

\subsection{Evolution of Dust Extinction at $0\lesssim z\lesssim 4$}
\label{sec:dust_evolution}

We investigate the evolution of dust extinction.
Figure \ref{fig:extinction_redshiftdepend} shows the histogram
of $E(B-V)$ for the whole LBGs at $z=4$ ($i'<25.5$, i.e., $M\lesssim M^*$), 
together with those for local starburst galaxies (local SBs; 
\citealt{meurer1999}) and
LBGs at $z=3$ (\citealt{adelberger2000}; $R\leq 25.5$, i.e.,
$M\lesssim M^*+1$).
The $E(B-V)$ values for local galaxies and $z=3$ LBGs
have also been derived from UV slopes using
the extinction-$\beta$ relation of 
\citet{meurer1999} and the dust extinction curve 
of \citet{calzetti2000}. The extinction-$\beta$
relation used in these papers 
is similar to ours which is shown in Figure \ref{fig:comp_meurer1999}
with only a small difference of $\Delta A_{\rm 1600}\lesssim 0.3$ 
corresponding to $\Delta E(B-V)\lesssim 0.03$. 

The mean dust extinction is
0.20, 0.15, and 0.15 for local SBs, $z=3$ LBGs, and $z=4$ LBGs, 
respectively.
If taken at face value, the mean dust extinction of LBGs at $z=3-4$
is lower than that of local starburst galaxies. 
%
%
%
However, this trend may be superficial.
The sample of local starbursts is based on a combination of
galaxy catalogs which are constructed by observations
at various wavelengths,
while the LBG samples are UV-continuum limited samples 
(section \ref{sec:photometric_lbg_galaxies}).
One cannot rule out the possibility that 
the observed trend of $E(B-V)$ is due to the selection effect 
that UV-continuum limited samples are likely to be biased 
toward dust-poor objects.
%
%
%
%
On the other hand, the sample selections of LBGs at $z=3$ and $z=4$
are quite similar. The very small difference in the mean extinction
between $z=3$ and $z=4$ LBGs, at most 
$\Delta E(B-V)<0.03$, indicates that
there is no evolution in dust extinction between $z=3$ and $4$.

\section{Evolution of UV-Luminosity Density}
\label{sec:evolution}

We calculate UV-luminosity densities of our LBGs
from the luminosity functions (LFs) 
derived in section \ref{sec:luminosity}.
First, we integrate the LFs (Figure \ref{fig:lumifun_comb_SFR})
down to 
the magnitudes of the faintest LBGs in our samples
(i.e., down to $M_{\rm 1700}=-19.8$, $-20.5$, $-20.5$
for $BRi$-LBGs, $Viz$-LBGs, and $Riz$-LBGs, respectively)
to obtain observed UV-luminosity densities, $\rho_{\rm UV}^{\rm obs}$,
the lower limits of the UV-luminosity density.
The total UV-luminosity density has to be larger than
$\rho_{\rm UV}^{\rm obs}$ by the contribution from galaxies fainter than
the limiting magnitudes.
In order to estimate the total UV-luminosity density, 
$\rho_{\rm UV}^{\rm total}$,
we extrapolate the LFs down to $L=0.1L^*$.
We also calculate the total UV-luminosity density
by extrapolating the LFs down to $L=0$ 
assuming $\alpha=-1.6$ which
are the upper limits of the UV-luminosity density.
To derive $\rho_{\rm UV}^{\rm total}$ down to $L=0$,
we use the analytic formula $\rho_{\rm UV}^{\rm total}=
L^* \phi^* \Gamma(\alpha+2)$.
Both $\rho_{\rm UV}^{\rm obs}$ and $\rho_{\rm UV}^{\rm total}$ 
for $BRi$-LBGs, $Viz$-LBGs, and $Riz$-LBGs
are summarized in Table \ref{tab:lumifun_schechter}.
The upper limits of the UV-luminosity densities are shown
in parentheses in Table \ref{tab:lumifun_schechter}.
In Table \ref{tab:lumifun_schechter}, we also show
the results of Lyman $\alpha$ emitters at $z=4.86\pm 0.03$
which are similarly calculated from the best-fit UV-luminosity
function obtained in SDS II.

%
%
%
We discuss the evolution of star-formation rate 
(section \ref{sec:evolution_star}) and 
the reionization of the universe 
(section \ref{sec:evolution_contribution}) using 
the $\rho_{\rm UV}^{\rm total}$ estimates.
We do not use $\rho_{\rm UV}^{\rm obs}$ for our discussion,
since our data are still 
considerably shallow and thus objects fainter than 
the limiting magnitudes will certainly 
dominate the total luminosity density.
The $\rho_{\rm UV}^{\rm total}$ values for 
$\alpha=-2.2$ and $-1.6$ gives an estimate of 
how much the $\rho_{\rm UV}^{\rm total}$ values 
are dependent on the choice of the faint-end slope.
We find in Table \ref{tab:lumifun_schechter} that 
for each sample, the $\rho_{\rm UV}^{\rm total}$ values 
down to $L=0.1L^*$ for the two $\alpha$ agree with each other 
within the errors.
%
%
%
%
%
%
Furthermore, the trend that $\rho_{\rm UV}^{\rm total}$ decreases 
slightly from $z=4$ to 5 is consistently seen 
for both $\alpha$ values.
Thus, the error in $\rho_{\rm UV}^{\rm total}$ 
due to the change in $\alpha$ seems to be modest 
over a reasonable range of $\alpha$.
%
%
%
%
%
%
Although the choise of the faint-end slope
does not affect much on the results, the estimated
total luminosity density is based on the large extrapolation
to the observed luminosity density. We show how much this extrapolation
affects to the results of
the following section (section \ref{sec:evolution_star_evolution}).
%
%
%
Since we obtain the upper limits of luminosity densities
for the case of $\alpha=-1.6$, we adopt the luminosity densities
from the Schechter parameters with $\alpha=-1.6$ (fixed) for
the following sections.

\subsection{Star-Formation Rate Density}
\label{sec:evolution_star}

\subsubsection{Evolution of Star-formation Rate Density Based on $L_{\rm 2000}$}
\label{sec:evolution_star_evolution}

We calculate the cosmic star-formation rate (SFR) from the
UV-luminosity density, $\rho_{\rm UV}^{\rm total}$. 
We use the relation between the UV luminosity
and the star formation rate given by \citet{madau1998}:
\begin{equation}
{\rm SFR(M_\odot yr^{-1})} = 
L_{UV}{\rm (erg\ s^{-1} Hz^{-1})}/(8\times 10^{27}),
\label{eq:starformationrate}
\end{equation}
where $L_{\rm UV}$ is the UV luminosity
measured in 1500\AA$-$2800\AA.
\footnote{
The conversion factor in eq. (\ref{eq:starformationrate})
is $8.0\times 10^{27}$ for the luminosity
at $1500$\AA\hspace{2pt} and $7.9\times 10^{27}$ 
for the luminosity at $2800$\AA\ \citep{madau1998}.
}
This relation assumes that galaxies have 
the Salpeter IMF with solar metallicity.
This conversion is insensitive to the difference in the
star formation history especially for the far-UV 
luminosity at $\lesssim 2000$\AA, since UV fluxes are
produced by massive OB stars whose lifetimes are
$\lesssim 2\times10^7$ yr \citep{madau1998}.
Figure \ref{fig:madauplot} shows the cosmic SFRs of $z=4-5$ LBGs
in our sample, together with those of galaxies at 
$z=0$ \citep{sullivan2000},
$z=0-1$ \citep{cowie1996},
and $z=3-4$ \citep{steidel1999}.
These cosmic SFRs (and UV-luminosity densities) 
have been corrected for the same amount of
dust correction, $E(B-V)=0.15$, since there is no
significant change in the observed $E(B-V)$ value over
$z=0-4.5$ (section \ref{sec:dust_evolution}).
%
%
%
We also show the cosmic SFRs at $z=0$, $0.2$, $0.9$, and $1.3$
estimated from H$\alpha$ luminosity density 
given by \citet{gallego1995}, \citet{tresse1998},
\citet{glazebrook1999}, and \citet{yan1999}, respectively. 
%
%
%
%
Note that the cosmic SFRs
estimated from the UV-luminosity density with dust correction
are comparable to those calculated from H$\alpha$ luminosity density.
Figure \ref{fig:madauplot} indicates that 
there is no significant change, or possibly a slight decline,
in the cosmic SFR
from $z\simeq 1$ up to $z\simeq 5$.
%
%
%
The possible decline from $z=4$ to $z=5$ is due to
the decrease in luminosity density found in the
previous section.
%
%
%

%
%
%
In Figure \ref{fig:madauplot}, we also plot
the star-formation rate
calculated from $\rho_{\rm UV}^{\rm obs}$ (open circles).
Since $\rho_{\rm UV}^{\rm obs}$
is the UV-luminosity density contributed by the bright portion of
LFs for which data exist, these are robust lower limits 
of the star-formation rate. 
We find from our data at $z=4$ that the cosmic SFR is almost
constant from $z\simeq 1$ to $z=4.5$ 
even when we use $\rho_{\rm UV}^{\rm obs}$ instead of 
$\rho_{\rm UV}^{\rm tot}$. 
It should be noted that the lower limit at $z=4$ 
obtained by Steidel et al. (1999) is much lower than 
ours based on $\rho_{\rm UV}^{\rm obs}$ due to 
their shallow data. 
Our data have improved largely the robust lower limit of the cosmic 
SFR at $z=4$.
On the other hand, the lower limit derived for our $z=5$ LBGs 
is not high enough to reject a large decline
of the cosmic SFR at $z=5$. 
However the true cosmic SFR at $z=5$ is presumably high 
as the estimate from $\rho_{\rm UV}^{\rm tot}$ indicates, 
since we find at $z\lesssim 1$ 
a good agreement between the cosmic SFRs derived from 
total H$\alpha$ luminosity densities  
and those based on UV-luminosity densities which are 
estimated from extrapolated LFs with dust-extinction correction.
%
%
%
%
%
%
We fit the cosmic SFR data including those based on
H$\alpha$ luminosity density by the analytic function of redshift
given in \citet{cole2001}: 
cosmic $SFR=(a+bz)/(1+(z/c)^d) h_{70}^{-2} M_\odot/$yr$/$Mpc$^3$,
and obtain $a=0.0039$, $b=0.13$, $c=1.6$, and $d=1.8$.
%
%
%
%
%
We show the best-fit function with errors 
by the shaded region in Figure \ref{fig:madauplot}.

\subsubsection{Stellar Mass Assembly History}
\label{sec:evolution_star_stellar}

We estimate the stellar mass density accumulated
from $z=6$, using the best-fit function of cosmic SFR 
obtained in section \ref{sec:evolution_star_evolution}.
We show the stellar mass density 
calculated by integrating the cosmic SFR over cosmic time
as a function of redshift by the shaded region
in the bottom panel of Figure \ref{fig:madauplot},
together with those measured directly from 
the stellar mass function of galaxies 
at $z=0-3$ \citep{cole2001,cohen2002,dickinson2003}.
Since the cosmic SFRs at high-$z$ are calculated from an
extrapolation of luminosity function, they may overestimate
the real cosmic SFRs, if the true
luminosity functions have flatter slopes.
Thus, we also estimate the lower-limits of stellar mass density
from the observed luminosity densities (open symbols
in the top panel of Figure \ref{fig:madauplot}).
The lower-limit values are connected by the dashed line
in Figure \ref{fig:madauplot}.

At $z=0-1$, the stellar mass densities derived directly from
the stellar mass functions are consistent with those calculated
from the cosmic SFRs (both the total densities, $\rho_{\rm UV}^{\rm total}$,
and the lower limits, $\rho_{\rm UV}^{\rm obs}$) 
within the uncertainties. 
At $z=1-3$, however, the stellar mass densities 
based on the stellar mass functions
are as low as the lower limits calculated from 
$\rho_{\rm UV}^{\rm obs}$ and about a factor of
three lower than the total densities calculated from
$\rho_{\rm UV}^{\rm total}$. 
There are at least four possible reasons for this discrepancy.
First, the stellar mass densities obtained by \citet{dickinson2003} 
may suffer from a large cosmic variance because they are based on 
data of the HDF-N, a very small patch of the sky.
Second, the stellar population synthesis models used to 
derive the cosmic SFRs and the stellar mass functions 
may not be appropriate for high-$z$ galaxies.
For example,
both our analysis and \citet{dickinson2003}'s adopt 
the Salpeter IMF. 
However, it is possible that
high-$z$ galaxies have a different IMF (e.g., top-heavy IMF).
Third, \citet{dickinson2003} assume 
a constant mass-to-luminosity ratio
for faint galaxies whose stellar masses
are not measured by fitting their stellar synthesis
models. If, however, the actual mass-to-luminosity ratio 
increases with decreasing luminosity, the stellar mass densities 
obtained by \citet{dickinson2003} are underestimates 
of the true values.
Fourth, while we assume that the dust extinction of LBGs 
is constant with luminosity, LBGs fainter than the detection limits 
may have smaller extinction. 
Such a decrease in $E(B-V)$ could change largely the total 
cosmic SFR, since the contribution to the cosmic SFR 
from LBGs fainter than the detection limits is significant.

\subsection{Contribution to the Reionization of the Universe}
\label{sec:evolution_contribution}

%
%
%
The IGM has been ionized since, at least,
$z=6$ \citep{becker2001,fan2002}. 
%
%
%
Since a large number of ionized hydrogens recombine 
in a relatively short time scale,
\footnote{
The recombination time scale depends on the density of
IGM and thus on redshift. The recombination time
scale at $z=3$ is estimated to be $\approx 300$ Myr 
\citep{madau1999}.
}
ionizing photons have to be supplied by objects at each epoch
to keep the IGM ionized.
\citet{madau1999} give a formula to calculate
the critical rate of ionizing photons, $\dot{N}_{\rm ion}^{\rm cr}$,
required to maintain the ionization of IGM.
The original formula is given in Einstein-de Sitter cosmology,
and we rewrite their formula, which can be applied
to our $\Lambda$-cosmology ($h=0.7$,$\Omega_m=0.3$,
$\Omega_\Lambda=0.7$) with an acceptable accuracy
at $3\lesssim z\lesssim 6$, as: 
\begin{equation}
\dot{N}_{\rm ion}^{\rm cr} =
(10^{51.0}\ {\rm s^{-1} Mpc^{-3}}) C_{30} \left(\frac{1+z}{6}\right)^3 
\left(\frac{\Omega_m h^2}{0.02}\right)^2,
\label{eq:critical_rate}
\end{equation}
where $C_{30}$ is the ionized hydrogen clumping factor
normalized by 30. The fiducial value
for this clumping factor is $C_{30}=1$.
The main uncertainty in this critical rate
is originated from this clumping factor, 
which is estimated to be of order $\pm 0.2$ in 
the log \citep{madau1999}.
We plot the critical rate as a function of redshift 
in Figure \ref{fig:reionize_Ndot_z}, together
with the emission rate of ionizing photons from QSOs
shown in \citet{madau1999}. Figure \ref{fig:reionize_Ndot_z}
indicates that the QSOs' production rate of ionizing photons,
$\dot{N}_{\rm ion}{\rm (QSO)}$,
is less than the critical rate at $z\gtrsim 3.6$.

Since the number density of low-luminosity AGNs 
at high-$z$ ($z\sim 3$) is much lower than that of
LBGs at similar redshifts (3\%; \citealt{steidel2002}),
the deficit of ionizing photons should be supplied from
massive stars in galaxies.
We estimate the emission rate of ionizing photons
per unit volume from galaxies, $\dot{N}_{\rm ion}{\rm (GAL)}$,
at $z=4$ and $5$. The emission rate is related
to the cosmic SFR, i.e.,
the star-formation rate density (SFRD):
\begin{equation}
\dot{N}_{\rm ion}{\rm (GAL)}=C f_{\rm esc} 
{\rm SFRD [M_\odot yr^{-1}Mpc^{-3}]}\ \ \ 
{\rm Number s^{-1} Mpc^{-3}},
\label{eq:emission_rate}
\end{equation}
where $C$ is a conversion factor, $f_{\rm esc}$ is
the escape fraction of ionizing photons,
and SFRD is the star-formation rate of galaxies
per unit volume \citep{madau1999}. 
\citet{madau1999} estimate
$C=10^{53.1}$. We use eq. (\ref{eq:starformationrate})
to calculate the star-formation rate.

Among the above parameters,
the escape fraction of ionizing photons is unknown. Here we
give an constraint on the escape fraction 
for LBGs at $z=4$ and $z=5$ using our data as follows.
Since the IGM at $z\lesssim6$ is ionized \citep{becker2001}, 
the sum of ionizing photons from QSOs and galaxies should
exceed the critical rate, at least, at $z\lesssim 6$:
\begin{equation} 
\dot{N}_{\rm ion}^{\rm cr}<
\dot{N}_{\rm ion}{\rm (GAL)}+\dot{N}_{\rm ion}{\rm (QSO)}.
\label{eq:ionization_constraint}
\end{equation}
First, we consider the escape fraction of our LBGs at $z=5$.
We estimate the critical rate using eq. (\ref{eq:critical_rate})
to be $\dot{N}_{\rm ion}^{\rm cr}=8.8^{+5.1}_{-3.2} \times 10^{50}$,
and we find $\dot{N}_{\rm ion}{\rm (QSO)}=3.0\times 10^{50}$
\citep{madau1999}. On the other hand, the number density of
ionizing photons from galaxies is calculated to be
$\dot{N}_{\rm ion}{\rm (GAL)}=4.4 \pm 1.9 \times 10^{51} f_{\rm esc}$.
Substituting these values for eq. (\ref{eq:ionization_constraint}),
we find $f_{\rm esc}>0.13^{+0.13}_{-0.09}$ for LBGs at $z=5$.
Similarly we obtain $f_{\rm esc}>0.02^{+0.05}_{-0.02}$ 
for LBGs at $z=4$. Note that the errors in $f_{\rm esc}$
include the uncertainty in $C_{30}$, $\pm 0.2$ dex. Thus,
we place a moderately significant constraint on the escape fraction for
LBGs at $z=5$, but not for LBGs at $z=4$.
Throughout the above discussion, we use the upper limits of
the luminosity density (values in parentheses in Table 
\ref{tab:lumifun_schechter})
for LBGs at $z=4$ and $5$,
which are obtained by integrating the LFs down to $L=0$.
Since $f_{\rm esc}$ decreases with the luminosity density,
the $f_{\rm esc}$ values calculated from these
luminosity densities are regarded as conservative lower limits.
We conclude that the escape fraction should be
$f_{\rm esc}\gtrsim 0.13$ for LBGs at $z=5$.
We plot in Figure \ref{fig:reionize_Ndot_z}
the sum of the number densities of ionizing photons
from QSOs and galaxies at $z=3-5$, 
assuming the lower-limit of the escape fraction
to be $f_{\rm esc}=0.13$.

\section{Discussion}
\label{sec:discussions}

\subsection{Luminosity Functions and Dust Extinction of LBGs 
at $z=4-5$}
\label{sec:discussions_luminosity}

 Luminosity functions (LFs) of LBGs show that
the number of bright galaxies significantly decreases
from $z=4$ to $z=5$, but little from $z=3$ to $z=4$.
We find that the slope of the LF may steepen from $z=3$ to $z=4$.
Our findings indicate that most of the bright ($M_{\rm 1700}\lesssim -22$)
galaxies appear between $z=4$ and $5$,
while faint ($M_{\rm 1700}\gtrsim -22$) galaxies dominate in number density
at $z\gtrsim4$. 
We find in Figure \ref{fig:lumifun_comb_SFR} that while
our LF at $z=4$ agrees well with that derived by \citet{steidel1999},
our LF at $z=5$ is different from that
obtained by \citet{iwata2003}
who claim that the LF at $z=5$ is similar to
that at $z=3$.
Thus there is a discrepancy between our and their
findings. We examine the cause of this discrepancy.
First of all, the field variance may be large
for such bright LBGs, since the number of detected bright LBGs
is as small as $\sim 20$.
However, we find an excellent consistency between the
LFs derived from the SDF and the SXDF. Thus the
field variance is probably not the main reason for
this discrepancy.
Second, it is possible that
the selection criteria of \citet{iwata2003}
would take a large number of contaminants
in their LBG sample as discussed 
in section \ref{sec:photometric_lbg_definitions},
resulting in the larger number density of bright
LBGs than by ours.

We estimate the star-formation rate of 
bright LBGs with $M_{\rm 1700}\lesssim -22$
with eq. (\ref{eq:starformationrate} to be
$SFR_{\rm raw}\gtrsim 30 h_{70}^{-2} M_\odot$yr$^{-1}$
(see the upper abscissa axis of Figure \ref{fig:lumifun_comb_SFR}
for the correspondence between $M_{\rm 1700}$ and
$SFR_{\rm raw}$). Since the UV luminosity of 
these LBGs has dust extinction
of $E(B-V)\simeq 0.15$,
the extinction corrected SFR is about 
a factor of 4 larger 
than $SFR_{\rm raw}$ (section \ref{sec:dust_dust}).
Thus LBGs with $M_{\rm 1700}\lesssim -22$
have intrinsic $SFR$ of $\gtrsim 100 h_{70}^{-2} M_\odot$yr$^{-1}$.
The deficit of $M_{\rm 1700}\lesssim -22$
LBGs at $z=5$ indicates that the number density of galaxies with
high-star formation rate of $\gtrsim 100 h_{70}^{-2} M_\odot$yr$^{-1}$
drops from $z=4$ to $z=5$.

Our LFs are derived for UV-luminosity or, equivalently, 
star-formation rate.
In general, galaxies with larger sizes and higher 
star-formation efficiencies have higher star-formation rates.
Thus, our findings imply that large galaxies are formed
by subsequent mergers of small galaxies
and/or that star-formation efficiency increases from $z=5$ to
$z=3$. The former interpretation supports the picture
of hierarchical clustering (e.g., \citealt{baugh1998,
kauffmann1999,weinberg2002}) in which galaxies experience 
a number of mergers. On the other hand,
the latter interpretation is not consistent with predictions
from numerical simulations of hierarchical clustering. 
Since the cooling efficiency
increases with the density of gas in dark halos,
hot gas cools more efficiently
at higher redshifts,
resulting in a higher star-formation efficiency at a higher redshift
\citep{hernquist2003}, which is opposite to the observed evolution.
Thus, a decrease in
the number density of dark halos 
predicted by the former scenario
should explain the
observed evolution of the LF at bright magnitudes.
We calculate the cumulative number density
of massive dark halos down to $10^{12}M_\odot$, where
LBGs with $M_{\rm 1700}\lesssim -20.5$ are expected to reside 
(see the companion paper, SDS VI),
and find that the number density of 
these massive halos is 44\% (at $z=4$) 
and 16\% (at $z=5$) of that at $z=3$. This decrease
is roughly consistent with the observed decrease 
of bright LBGs in the number density
from $z=3$ to $z=5$.
%
%

The slope of the LF at $z=4$ is estimated to be $\alpha=-2.2\pm 0.2$.
Since the faintest LBGs in our sample are as bright as $M^*+1$, 
the estimated slope
should have large systematic errors. 
If taken at face value,
the faint-end slope of the LF at $z=4$ is steeper than
that at $z=3$ ($\alpha=-1.6$; \citealt{steidel1999}).
\citet{yan2003} report that the faint-end slope of 
the LF for LBGs at $z\simeq 6$ may be as steep as $\alpha\sim -2$
from their deep HST/ACS data. Thus, the slope of the LF
for LBGs may steepen at $z\gtrsim 4$. However, 
these results might 
conflict with model predictions. 
The number density of faint galaxies is predicted to
decrease right after
the reionization ($z\simeq 6$; \citealt{becker2001}),
since the reionization increases
the Jeans mass of the IGM and thus
the minimum mass of forming
galaxies \citep{gnedin1997,miralda-escude1998}.

The average extinction of LBGs at $z=4$,
$E(B-V)=0.15\pm 0.03$, is the same as 
that of LBGs at $z=3$ derived by \citet{adelberger2000}
from their large sample of $z=3$ LBGs.
Furthermore, the dependence of extinction on intrinsic luminosity 
for $z=4$ LBGs is similar to that for $z=3$ LBGs.
This means that the average dust properties are the same between
$z=3$ LBGs and $z=4$ LBGs. Interestingly, \cite{ferguson2002}
find that most of the LBGs seen at $z=3$ started star formation
after $z=4$, i.e., LBGs at $z=3$ are not descendants of LBGs
at $z=4$. Thus, LBGs at $z=3$ and $z=4$ may be similarly
young, and so have a similar amount of dust.

\subsection{UV-Luminosity Density and Escape Fraction}
\label{sec:discussions_uv}

     %

We compare in Figure \ref{fig:madauplot} 
the luminosity densities of LBGs at $z=4-5$ in our samples 
with those at $z=3$ and 4 which are calculated from the LFs 
given in \citet{steidel1999}.
Figure \ref{fig:madauplot} shows that the UV-luminosity density
of LBGs does not change significantly from $z=3$ to $z=5$.
The luminosity density of LBGs is
$\rho_{\rm UV}^{\rm total}=$
$1.9\pm0.2\times 10^{26}$, 
$2.0\pm0.2\times 10^{26}$, and
$1.6\pm0.7\times 10^{26}$ erg s$^{-1}$ Hz$^{-1}$ Mpc$^{-3}$
(see Table \ref{tab:lumifun_schechter})
for $z=3$, $z=4$, and $z=5$, where the value at $z=4$ is the
mean of our measurement and \citeauthor{steidel1999}'s.
Thus the ratio of the UV-luminosity density at 
$z=4$ and 5 to that at $z=3$
is
$\rho_{\rm UV}(z=4)/\rho_{\rm UV}(z=3)=1.0\pm0.2$ and
$\rho_{\rm UV}(z=5)/\rho_{\rm UV}(z=3)=0.8\pm0.4$.
The total UV-luminosity density may slightly decrease toward $z=5$,
but the amount of the decrease is 20\% at most.
Although the number density of bright 
LBGs ($M_{\rm 1700}\lesssim -22$)
significantly decreases toward $z=5$ (Figure \ref{fig:lumifun_comb_SFR}), 
the total UV-luminosity density does not change largely.
This is because the total UV-luminosity density is
mainly contributed by LBGs fainter than
$M_{\rm 1700}\sim -22$. 

The luminosity density of LAEs at $z=4.86$ is calculated to be
$9.6\times 10^{25}$ erg s$^{-1}$ Hz$^{-1}$ Mpc$^{-3}$
with the UV-luminosity
function obtained by SDS II
in the same manner as for LBGs (Table \ref{tab:lumifun_schechter}).
The ratio of the UV-luminosity density of LAEs to that of LBGs is 
$\rho_{\rm UV}({\rm LAE})/\rho_{\rm UV}({\rm LBG})\simeq 0.6$
at $z\simeq 5$. Since LBG samples are 
UV-luminosity limited samples 
(section \ref{sec:photometric_lbg_galaxies}),
the UV-luminosity density derived from LBGs' LF 
represents the UV-luminosity density of the whole galaxy population 
(if an appropriate extrapolation of the LF down to a very faint 
luminosity is made).
Thus, about 60\% of the cosmic 
UV-luminosity density (or cosmic star-formation rate) 
at $z\sim 5$ is contributed by galaxies 
identified as LAEs. 


Figure \ref{fig:madauplot} shows that the cosmic
SFR is constant from $z\sim1$ to $z\sim 5$.
This result agrees with that obtained by \citet{iwata2003},
although the bright end of the luminosity function
is significantly different between ours and
\citeauthor{iwata2003}'s (\citeyear{iwata2003}).
This coincidence indicates that faint LBGs 
($SFR\sim 1 h_{70}^{-2} M_\odot$yr$^{-1}$) contribute
much more to the cosmic SFR than bright LBGs
($SFR\gtrsim 100 h_{70}^{-2} M_\odot$yr$^{-1}$).
The constant cosmic SFR from $z=1-5$ is consistent with predictions of
numerical simulations.
\citet{ascasibar2002} have found from numerical simulations
that the cosmic star-formation rate shows almost no drop 
over $2<z<5$, and that the star-formation is a gradual
process with no characteristic epoch.
\citet{nagamine2000} have predicted from hydrodynamical
simulations that 
the cosmic SFR shows a moderate plateau
between $z=1$ and $z=3$ and a gradual decrease beyond $z=3$
up to $z=5$ by $\sim 0.4$ dex. 
The cosmic SFR at $z=3-5$ is about 5
times larger than that at $z=0$ (top panel of
Figure \ref{fig:madauplot}).
The star-formation is very active at these
high redshifts, but the accumulated stellar mass density 
(bottom panel of Figure \ref{fig:madauplot}) at $z=3$ is just
$1/10$ of the present-day stellar mass density.
Since the cosmic time between $z=5$ and $3$ is much shorter than
that between $z=3$ and $0$ (1Gyr versus 10Gyr),
the majority of stars are produced not at these high redshifts
($z=3-5$) but at lower redshifts ($z<3$).


We calculate the number density of ionizing photons
contributed by LBGs (section \ref{sec:evolution_contribution}). 
We give a lower-limit for the escape fraction of 
ionizing photons for LBGs at $z=4.7$
($f_{\rm esc}\gtrsim 0.13$).
%
%
%
It should be noted here that this $f_{\rm esc}$ value is inferred 
from a combination of the estimated UV-luminosity for LBGs 
and a model of ionization. 
In this sense, our method has intrinsically a large ambiguity, 
since most of the $f_{\rm esc}$ values given in the literature 
are based on direct measurements of the Lyman continuum in spectra 
(see below). 
Nevertheless, our results are useful, giving a new, independent 
constraint on $f_{\rm esc}$.
%
%
%

\citet{steidel2001} have found that the average spectrum of LBGs 
at $z\simeq 3.4$ has a Lyman continuum and that 
the flux ratio between the UV continuum at 1500\AA\ 
and the Lyman continuum at 900\AA\ is $f_{900}/f_{1500}=1/4.6$
after correction for the IGM absorption.
This flux ratio corresponds to an escape fraction,
$f_{\rm esc}\sim 3\times (1/4.6) = 0.65$, where
the factor of 3 comes from the assumed shape of 
the intrinsic spectrum (see \citealt{giallongo2002}).
On the other hand, \citet{giallongo2002} have found no Lyman-continuum
flux in their two LBGs at $z=3$, and they set an upper limit
for the escape fraction of $f_{\rm esc}<0.16$.
%
%
%
The measurement of \citet{giallongo2002} is not consistent with
that of \citet{steidel2001}. \citet{giallongo2002} discuss
that this inconsistency may be due to differences 
in sample selection.
In either case, our estimate, $f_{\rm esc}\gtrsim 0.13$, is 
not seriously conflict with these two previous measurements, 
if the escape fraction does not change from $z=5$ to $z=3$.
The escape fraction of LBGs is probably not smaller than $\sim 10\%$. 

We compare these $f_{\rm esc}$ values with those 
of the present-day galaxies.
%
%
%
%
%
%
%
%
\citet{leitherer1995b}
have given an upper limit for four nearby starbursts 
using Far-UV spectra.
The values they have obtained 
are $f_{\rm esc}\leq 0.0095$, $0.017$,
$0.048$, and $0.15$. 
\citet{hurwitz1997} re-analyzed the data
of \citet{leitherer1995b} to obtain higher values: 
$f_{\rm esc}\leq 0.032$, $0.052$,
$0.11$, and $0.57$.
\citet{tumlinson1999} have found $f_{\rm esc}\leq 0.02$ for
NGC 3067.
The escape fraction of the Milky Way Galaxy has been 
estimated to be $f_{\rm esc}\sim 0.06$ by \citet{bland-hawthorn1999}.
The average $f_{\rm esc}$ seems to be $f_{\rm esc}\lesssim 0.1$ 
for present-day galaxies. This implies that 
the escape fraction of LBGs at $z=3-5$ 
is larger than that of star-forming galaxies at $z=0$. 

%
%


\section{Conclusions}
\label{sec:conclusions}

We have made large samples of 2,600
Lyman Break Galaxies (LBGs) at $z=3.5-5.2$ detected
in deep ($i'\simeq 27$) and wide-field (1,200 arcmin$^2$)
data of the
Subaru Deep Field (SDF) and the Subaru/XMM Deep Field (SXDF),
and have studied their photometric properties.
The major findings of our study are summarized as follows.\\

\begin{figure}
\epsscale{1.0}
\plotone{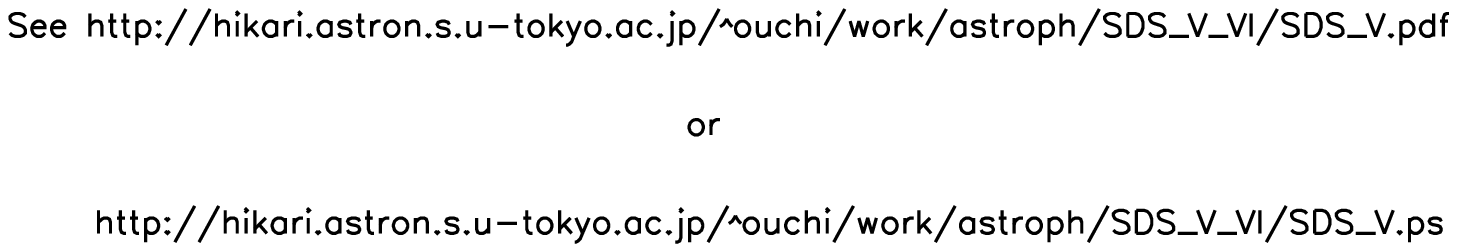}
\caption{
Redshift distribution of galaxies
in \citep{furusawa2000}'s HDF-N photometric redshift catalog.
In all panels, open histograms are for all
galaxies in the catalog.
Filled histograms are for
galaxies which satisfy at least one of our
$BRi$-, $Viz$-, or $Riz$-LBG selections (top panel),
$BRi$-LBG selection (second panel),
$Viz$-LBG selection (third panel), and 
$Riz$-LBG selection (bottom panel), respectively.
The top panel indicates that 
at $z=3.9-5.1$, four out of the 47 galaxies 
escape from our LBG selections.
\label{fig:photzhist_comb}}
\end{figure}

\begin{figure}
\epsscale{1.0}
\plotone{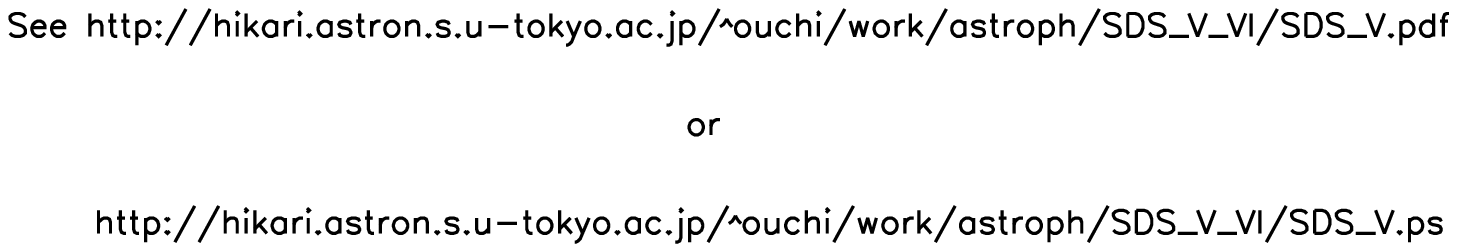}
\caption{
Histogram of $E(B-V)$ of galaxies 
in \citep{furusawa2000}'s HDF-N photometric redshift catalog.
Open histogram shows all galaxies in the catalog.
Shaded histogram shows the distribution of four galaxies
which are not selected by either of
the LBG selection criteria.
The four galaxies have larger 
dust extinction than $E(B-V)=0.35$.
\label{fig:missingobj_nebv}}
\end{figure}

\begin{figure}
\epsscale{1.0}
\plotone{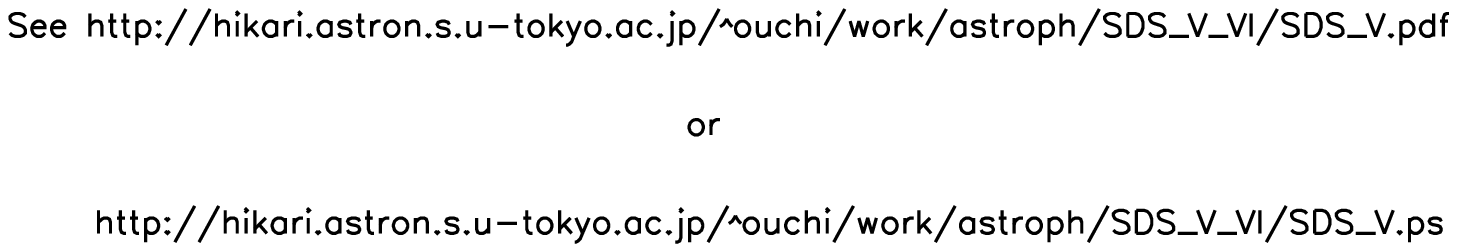}
\caption{
Luminosity functions (LFs) of LBGs at $z=4-5$.
LFs of $BRi$-, $Viz$-, and $Riz$-LBGs
are given in top panel, middle panel, and bottom panel,
respectively. In each panel, filled circles (open circles)
are the LFs derived from the SDF (SXDF) data. Solid lines
are the best-fit Schechter function whose parameters
are shown in Table \ref{tab:lumifun_schechter} (see text).
Dashed lines denote the LF of UV-selected galaxies in the
local universe \citep{sullivan2000}, while dotted lines
are the LF of LBGs at $z\simeq 3$ derived by \citet{steidel1999}.
In the top panel, the best-fit Schechter function of 
LBGs at $z=4$ \citep{steidel1999} is shown by the
dash-long dashed line down to $\sim -21$ mag.
\citet{steidel1999} measured the LF down to $\sim -21$ mag
with their wide-field
LBG survey and fitted the Schechter function with a 
fixed slope ($\alpha = -1.6$). 
The upper abscissa axis, $SFR_{\rm raw}$, 
indicates the star-formation rate without extinction correction, 
which is converted from $M_{\rm 1700}$ using 
eq. (\ref{eq:starformationrate}). 
The true (extinction-corrected) star-formation
rate is about a factor of 4 larger than
the raw rate. See text for details.
\label{fig:lumifun_comb_SFR}}
\end{figure}

\begin{figure}
\epsscale{1.0}
\plotone{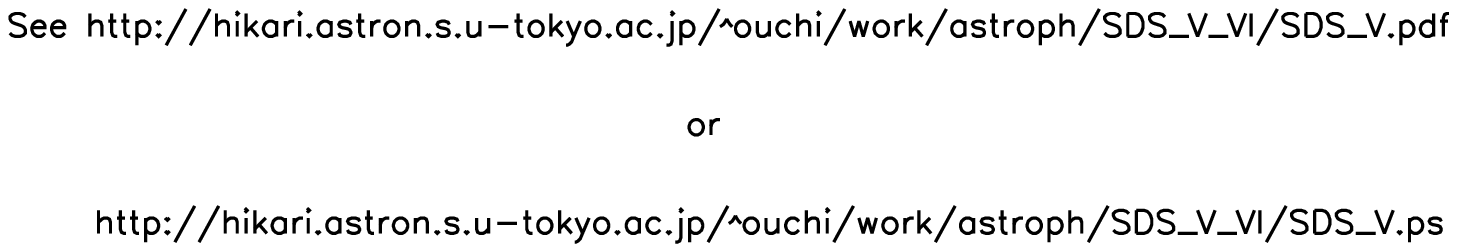}
\caption{
Internal absorption at $1600$\AA\ ($A_{\rm 1600}$) plotted
against $\beta$.
The solid line shows the
relation for the template spectrum (70 Myr age and $0.2Z_\odot$ metallicity).
Filled circles indicate 43 nearby star-forming galaxies, 
and the dashed line is a linear fit to the data given by
\citet{meurer1999}. 
The dotted line is for a model spectrum 
which has the same age as
the template spectrum (70Myr) but has a higher metallicity
of $1 Z_\odot$. The dot-dashed line is for a model
spectrum 
which has the same metallicity as the template spectrum ($0.2 Z_\odot$)
but has an older age of 90 Myr.
\label{fig:comp_meurer1999}}
\end{figure}

1. We find that our selection criteria for LBGs
can isolate about 90\% of all galaxies in a targetted
redshift range, if 
galaxies have sufficiently high $S/N$ ratios
(Figure \ref{fig:photzhist_comb}
in section \ref{sec:photometric_lbg_galaxies}). 
Thus, our LBG samples are regarded as nearly
UV-magnitude limited samples of high-$z$ galaxies.
The missed 10\% galaxies are galaxies
attenuated heavily by dust ($E(B-V)\gtrsim 0.4$).

2. We derive luminosity functions of LBGs at 
$\langle z\rangle =4.0$, $4.7$, and $4.9$
in section \ref{sec:luminosity}. 
We find no cosmic variance between the SDF and the SXDF.
Then comparing them with
that at $\langle z\rangle =3$ \citep{steidel1999},
we find that while the luminosity function of LBGs
does not show a large change over $z=3$ and $4$
as reported by \citet{steidel1999}, 
the number density of bright galaxies 
with $M_{\rm 1700}<-22$ (or galaxies with high star-formation
rate of $SFR\gtrsim 100 h_{70}^{-2} M_\odot$ yr$^{-1}$
with extinction correction)
decreases by an order of magnitude from $z=4$ to $5$.
We also find 
that the faint-end slope of the luminosity
function may be steeper at $z=4$ than at $z=3$.

\begin{figure}
\epsscale{1.0}
\plotone{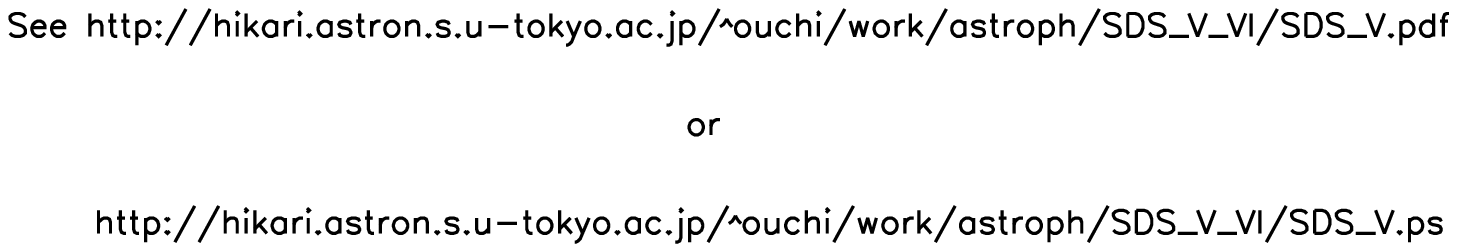}
\caption{
Redshift dependence of the 
$E(B-V)$-$\beta_{iz}$ relationship
calculated from the template model.
Solid line is the relation for LBGs at $z=4$,
and dashed and dot-dashed lines are for
$z=3.5$ and $z=4.5$, respectively. The difference
in the relation for LBGs at $z=3.5-4.5$ is less than 
$\pm 0.05$ mag in $E(B-V)$ for LBGs with $E(B-V)\leq 0.5$.
The upper and lower
dotted lines are the relation of LBGs at
$z=4.7$ and $z=5.2$, respectively.
\label{fig:redshift_beta}}
\end{figure}

\begin{figure}
\epsscale{0.9}
\plotone{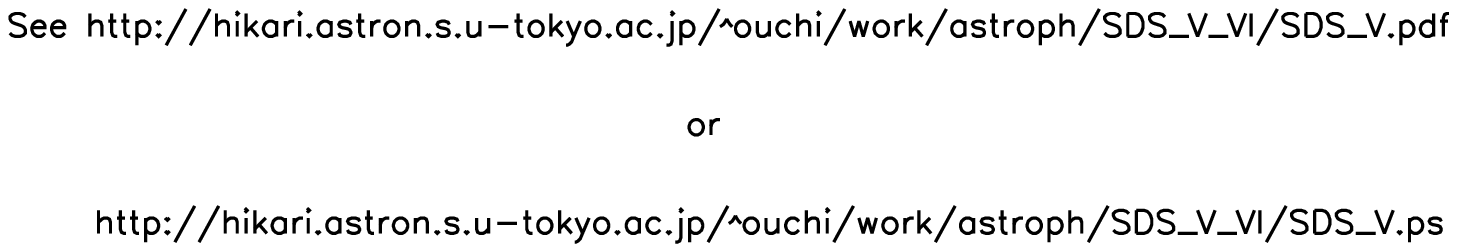}
\caption{
Histograms of $E(B-V)$ for four different magnitude
bins for $BRi$-LBGs in the SDF and SXDF. 
Top panel shows the completeness of our LBG detection
obtained by a simulations. In the top panel, the solid line
is the histogram of input objects for the simulation. 
Dotted, dashed, dot-dashed,
and long-dashed lines are the completeness of LBGs
with $i'=23.75$,$24.25$,$24.75$,and $25.25$, respectively.
Second top to bottom panels show
the number of objects in each magnitude bin
as a function of extinction. In these panels,
shaded histograms correspond to measurements corrected for
the completeness shown in the top panel, while open histograms
show raw numbers.
Arrows denote the mean value of extinction 
calculated from
the shaded histogram for each bin. Since the 
completeness for objects with 
large extinction values, $E(B-V)>0.55$, 
is less than 20\%, we do not include those objects in
the calculation of the mean value.
\label{fig:beta_BRi_dist_comb}}
\end{figure}

\begin{figure}
\epsscale{0.9}
\plotone{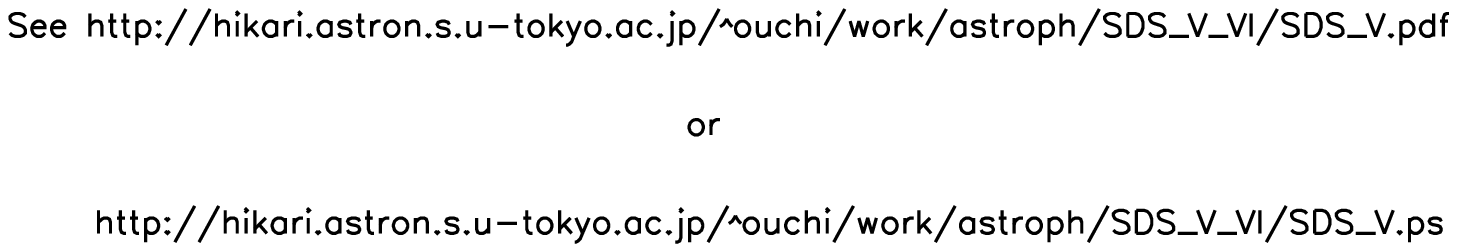}
\caption{
Histograms of $E(B-V)$ for galaxies at $z=0$, 
3, and 4.
Shaded histogram in the top panel shows 
the distribution of local starburst galaxies
derived from IUE data \citep{meurer1999}, while
the shaded histogram in the middle panel 
is for LBGs at $z=3$ \citep{adelberger2000}.
Dotted histogram in the bottom panel presents
our $BRi$-LBGs ($z=4$ LBGs) without correction for completeness.
The open histogram in each panel shows the distribution of
completeness-corrected $BRi$-LBGs
down to $i'=25$ (or $M\simeq M^*$). 
Arrows indicate the mean values of extinction 
for galaxies at each redshift, which are calculated from
the data over $0.0\leq E(B-V)\leq0.5$ for a fair comparison.
Contamination-corrected data are used for the calculation for
our $BRi$-LBGs and LBGs at $z=3$.
\label{fig:extinction_redshiftdepend}}
\end{figure}

\begin{figure}
\epsscale{0.7}
\plotone{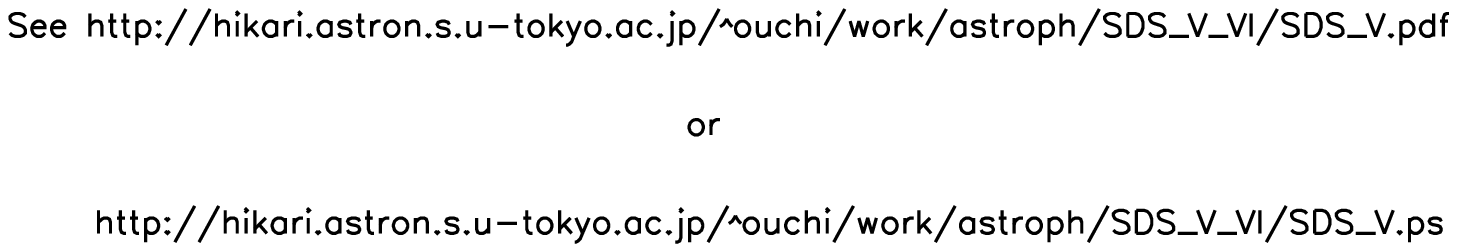}
\caption{
$Top\ panel:$
Cosmic star-formation rate (SFR)
as a function of redshift. The cosmic SFR
is calculated from the luminosity
density at $\simeq 2000$\AA, $\rho_{\rm 2000}^{\rm corr}$,
estimated from $\rho_{\rm UV}^{\rm tot}$ and $\rho_{\rm UV}^{\rm obs}$
which are corrected for dust extinction of $E(B-V)=0.15$ whose value
is indicated by a long arrow (see section \ref{sec:dust}).
Circles, squares, diamonds, and stars
are values for LBGs at $z=4-5$ (present study),
LBGs at $z=3-4$ \citep{steidel1999}, galaxies at 
$z\sim 1$ \citep{cowie1996}, and galaxies at $z\simeq 0$
\citep{sullivan2000}, respectively.
Filled symbols indicate the total cosmic SFRs
calculated by integrating the luminosity functions down to $L=0.1L^*$, 
and open symbols with arrows show lower limits, i.e., 
contributions only by actually detected galaxies.
Differences between filled symbols and open symbols are due to
the contributions from faint galaxies below the detection limits.
Open triangles plotted at $z>3$ indicate
the upper limits which are calculated
by integrating the luminosity functions down to $L=0$.
%
%
%
Pluses at $z=0$, $0.2$, $0.9$, and $1.3$ are cosmic SFRs 
derived from the H$\alpha$ luminosity density
(\citealt{gallego1995}, \citealt{tresse1998}, 
\citealt{glazebrook1999}, and \citealt{yan1999}, respectively). 
%
%
Note that the cosmic SFRs
estimated from $\rho_{\rm 2000}^{\rm corr}$ (filled symbols) 
are comparable to those calculated from the H$\alpha$ luminosity density
(pluses) at $z=0-1$. The shaded region shows approximated 
evolution of cosmic SFR obtained from fit of an analytic function 
(see text for details) to the data points.
$Bottom\ panel:$ Stellar mass density
as a function of redshift. The shaded region indicates 
the stellar mass density accumulated from $z=6$, which is
calculated from the cosmic SFR shown as the shaded region
in the top panel.
Dashed line shows the stellar mass density 
calculated from the cosmic SFR of 
observed galaxies alone (i.e., open symbols in the top panel).
Filled circles, squares, and
diamonds denote the stellar mass density derived 
from the stellar mass function by
\citet{cole2001}, \citet{cohen2002}, and 
\citet{dickinson2003}, respectively.
\label{fig:madauplot}}
\end{figure}

\begin{figure}
\epsscale{1.0}
\plotone{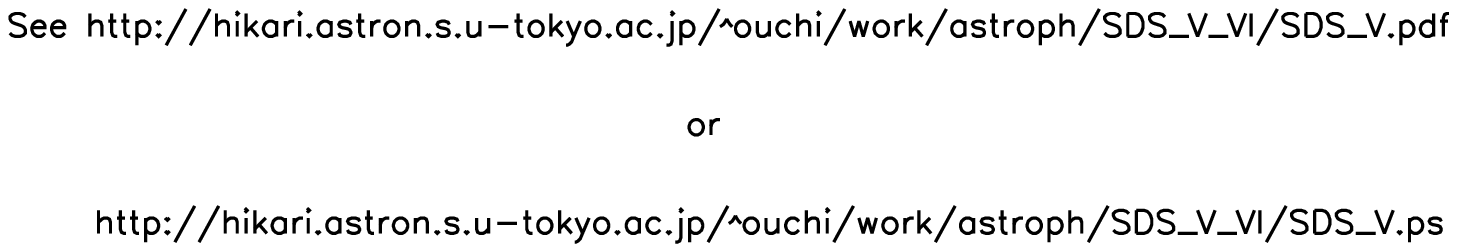}
\caption{
UV-ionizing photon density ($\dot{N}_{\rm ion}$) 
of the universe at $z=2-5$.
Solid line indicates $\dot{N}_{\rm ion}$ required
to maintain the ionization of the IGM 
predicted by \cite{madau1999}. Dashed line
is $\dot{N}_{\rm ion}$ contributed from
QSOs. 
Diamond and circles denote
the sum of the ionizing photons
from QSOs and galaxies, where the ionizing photons from
galaxies are estimated from LBGs
at $z=3$ (Steidel et al. 1999) and at $z=4$ and $5$ (present data) 
assuming $f_{\rm esc}=0.13$.
This figure visualizes the fact that
the average $f_{\rm esc}\gtrsim 0.13$ is required for galaxies
at $z=4.7$ in order to maintain ionization.
\label{fig:reionize_Ndot_z}}
\end{figure}

3. We estimate the dust extinction of LBGs at $z=4\pm 0.5$
from the UV-continuum slope measured
from $i'-z'$ color (section \ref{sec:dust}). 
We do not measure dust extinction of LBGs at $z>4.5$,
because $i'-z'$ measurements are significantly
affected by absorption of the IGM and by Ly$\alpha$ emission
(Figure \ref{fig:redshift_beta} of section \ref{sec:dust_uv}).
We find that LBGs with $M<M^* (\simeq -21)$
have $E(B-V)=0.15\pm0.03$ on average 
if completeness correction is made to the sample, 
and that the amount of extinction depends not on
apparent luminosity but on intrinsic luminosity.
%
We find no evolution in dust extinction
between $z=3$ and $4$. 

4. We calculate the UV-luminosity density at 1700\AA\ for 
our LBGs by integrating the luminosity functions
derived in section \ref{sec:luminosity_lbg}. Then we estimate
the cosmic star-formation rate at $z=4$ and $z=5$
from the UV-luminosity density,
and compare them with those at $z<4$
given by various authors.
We find that the UV-luminosity density at 1700\AA,
$\rho_{\rm UV}$, does not significantly
change from $z=3$ to $z=5$, i.e.,
$\rho_{\rm UV}(z=4)/\rho_{\rm UV}(z=3)=1.0\pm0.2$ and
$\rho_{\rm UV}(z=5)/\rho_{\rm UV}(z=3)=0.8\pm0.4$.
Comparing the UV-luminosity density of LBGs at $z=5$ with
that of Lyman $\alpha$ emitters (LAEs) at $z=4.9$
calculated from the data of SDS II,
we obtain 
$\rho_{\rm UV}({\rm LAE})/\rho_{\rm UV}({\rm LBG})\simeq 0.6$.
It implies that about 
a half ($\simeq 60$\%) of the star formation at $z\sim5$ occurs in LAEs.

5. We derive the cosmic star-formation rate (SFR)
at $z\sim4$ and 5 from $\rho_{\rm UV}$ of our LBGs
(section \ref{sec:evolution_star}) 
with correction for dust extinction of $E(B-V)=0.15$
obtained in section \ref{sec:dust_dust}.
Combining our measurements 
with those at $z\lesssim 3$ given in the literature, 
we find that the cosmic SFR is almost constant,
or shows a possible decline, from $z=3$ to $z=5$.
We then estimate the stellar mass density at $z\lesssim 5$ 
by integrating the cosmic SFR over time, 
and find that at $z\sim 1-3$ 
the stellar mass density based on the cosmic SFR 
exceeds that derived directly from the stellar mass function 
by a factor of 3, while the two estimates agree
at $z \lesssim 1$.

6. We estimate the production rate of ionizing photons for LBGs 
from $\rho_{\rm UV}$ using the model proposed by
\citet{madau1999} (section \ref{sec:evolution_contribution}).
We find that more than $\simeq 13$\% of
ionizing photons produced by massive stars should escape
from LBGs at $z\simeq 5$ (i.e., $f_{\rm esc}\gtrsim 0.13$)
in order to keep the IGM ionized.


\acknowledgments
We would like to thank the Subaru Telescope staff
for their invaluable help in commissioning the Suprime-Cam
that made these difficult observations possible.
We thank the referee, Patrick McCarthy, for 
his constructive suggestions and comments
that improved this article.
M. Ouchi acknowledge support from the Japan Society for the
Promotion of Science (JSPS) through JSPS Research Fellowships
for Young Scientists.



%
%



\clearpage 

\clearpage 

\clearpage 

\clearpage

\clearpage 

\clearpage 

\clearpage 

\clearpage 

\clearpage 

\clearpage 

\clearpage 

\clearpage

\clearpage 

\clearpage 

\clearpage

\clearpage

\clearpage

\clearpage 

\clearpage 

\clearpage 

\clearpage 

\clearpage

\clearpage 

\clearpage

\begin{deluxetable}{cclccr}
\tabletypesize{\scriptsize}
\tablecaption{Journal of observations \label{tab:obs}}
\tablewidth{0pt}
\tablehead{
\colhead{Field Name} & 
\colhead{Band} & 
\colhead{Observed dates} &
\colhead{Area (arcmin)} &
\colhead{Total Exposure Time (min)} &
\colhead{$m_{\rm lim}$} 
}
\startdata
SDF  & $B$  & 2001 Apr 24-25, May 20       & 616 & 210 &27.8 \\
SDF  & $V$  & 2001 Apr 23, May 20          & 616 & 150 &27.3 \\
SDF  & $R$  & 2001 Mar 21-23                 & 543 &  90 &27.1 \\
SDF  & $i'$ & 2001 Apr 24, Mar 19, Jun 23-24 & 616 & 138 &26.9 \\
SDF  & $z'$ & 2001 May 19-20, Jun 25        & 616 &  81 &26.1 \\
SXDF & $B$  & 2000 Nov 24-25                 & 653 & 177 &27.6 \\
SXDF & $V$  & 2000 Nov 26-27                 & 653 &  84 &26.5 \\
SXDF & $R$  & 2000 Nov 22,24, 2001 Nov 16    & 653 & 118 &27.2 \\
SXDF & $i'$ & 2000 Nov 25                    & 653 &  45 &26.2 \\
SXDF & $z'$ & 2001 Oct 14,18,19              & 653 &  40 &25.7 

\enddata

\tablecomments{
These data were taken in 2000 and 2001 during the GTOs of Suprime-Cam.
Although deeper imaging (3-10 hours for each band)
was carried out for both fields in 2002 and 2003
by the Subaru Observatory key projects (e.g., \citealt{kodaira2003};
see section \ref{sec:introduction}),
the results presented in this paper are based on the GTO data only.
}

\end{deluxetable}

\clearpage

\begin{deluxetable}{ccccccccc}
\tabletypesize{\scriptsize}
\tablecaption{Lyman break galaxies with spectroscopic redshifts
\label{tab:lbgs_with_redshifts}}
\tablewidth{0pt}
\tablehead{
\colhead{Object Name} & 
\colhead{RA(J2000)} &
\colhead{Dec(J2000)} & 
\colhead{$z$} &
\colhead{$B$} &
\colhead{$V$} &
\colhead{$R$} &
\colhead{$i'$} &
\colhead{$z'$}  
}
\startdata
SDFJ132410.8+272758 & 13:24:10.8 & +27:27:58 & 3.845 & 27.79 & 26.21 & 25.09 & 25.00 & 24.97\\
SDFJ132413.3+274207 & 13:24:13.3 & +27:42:07 & 4.140 & 27.84 & 26.10 & 25.32 & 25.16 & 25.17\\
SDFJ132416.3+274355 & 13:24:16.3 & +27:43:55 & 4.250 &$>29.0$& 25.54 & 24.54 & 24.09 & 24.02\\
SDFJ132413.1+274116 & 13:24:13.1 & +27:41:16 & 4.270 &$>29.0$& 27.32 & 26.08 & 26.10 & 25.91\\
SDFJ132411.4+273016 & 13:24:11.4 & +27:30:16 & 4.600 &$>29.0$& 26.61 & 25.40 & 24.66 & 24.51\\
SDFJ132410.5+274254 & 13:24:10.5 & +27:42:54 & 4.865 &$>29.0$&$>28.5$& 27.21 & 25.96 & 25.98
\enddata

\tablecomments{
Magnitudes are $2''\phi$ aperture magnitudes.
}

\end{deluxetable}

\clearpage 

\begin{deluxetable}{cccrlc}
\tabletypesize{\scriptsize}
\tablecaption{Photometric Samples of Galaxies\label{tab:sample}}
\tablewidth{0pt}
\tablehead{
\colhead{Field Name} & 
\colhead{Sample Name} &
\colhead{Detection Band} & 
\colhead{Number} &
\colhead{Magnitude Range\tablenotemark{\ddag}} &
\colhead{Selection Criteria} 
}
\startdata
SDF  & $BRi$-LBG & $i'$ & 1,438 & $i'=23.5-26.5$ & eq. (\ref{eq:lbgselection_BRiLBG})\\
SDF  & $Viz$-LBG & $z'$ &   246 & $z'=24.0-26.0$ & eq. (\ref{eq:lbgselection_VizLBG})\\
SDF  & $Riz$-LBG & $z'$ &    68 & $z'=24.0-26.0$ & eq. (\ref{eq:lbgselection_RizLBG})\\
SXDF & $BRi$-LBG & $i'$ &   732 & $i'=23.5-26.0$ & eq. (\ref{eq:lbgselection_BRiLBG})\\
SXDF & $Viz$-LBG & $z'$ &   34\tablenotemark{\dagger} & $i'=23.5-25.5$ & eq. (\ref{eq:lbgselection_SXDFVizLBG})\\
SXDF & $Riz$-LBG & $z'$ &    38 & $i'=24.0-25.5$ & eq. (\ref{eq:lbgselection_RizLBG})
\enddata

\tablenotetext{\ddag}{$2''\phi$ aperture magnitudes.
}
\tablenotetext{\dagger}{$Viz$-LBGs of the SXDF are selected 
by the equation 
\ref{eq:lbgselection_SXDFVizLBG}.
}

\end{deluxetable}

\clearpage

\begin{deluxetable}{cccccccc}
\tabletypesize{\scriptsize}
\tablecaption{Summary of the luminosity functions.
\label{tab:lumifun_schechter}}
\tablewidth{0pt}
\setlength{\tabcolsep}{0.02in}
\tablehead{
\colhead{Sample Name} &
\colhead{$z$} & 
\colhead{$\phi^*$} &
\colhead{$M_{\rm1700}^*$} &
\colhead{$\alpha$} &
\colhead{$n^{\rm obs}$\tablenotemark{\dagger}} &
\colhead{$\rho_{\rm UV}^{\rm obs}$\tablenotemark{\dagger}} &
\colhead{$\rho_{\rm UV}^{\rm total}$\tablenotemark{\dagger\dagger}} \\
\colhead{} &
\colhead{} & 
\colhead{($h_{70}^{3}$Mpc$^{-3}$)} &
\colhead{(mag)} &
\colhead{} &
\colhead{($h_{70}^{3}$Mpc$^{-3}$)} &
\colhead{(erg s$^{-1}$ Hz$^{-1}$)} &
\colhead{(erg s$^{-1}$ Hz$^{-1}$)} 
}
\startdata
$BRi$-LBG & $4.0^{+0.5}_{-0.5}$ 
 & $1.2\pm0.2\times10^{-3}$ & $-21.0\pm0.1$ & $-2.2\pm0.2$ 
 & $2.0\pm0.3\times10^{-3}$
 & $1.2\pm0.2\times10^{26}$ & $2.9\pm0.4\times10^{26}$ \\
$Viz$-LBG & $4.7^{+0.5}_{-0.5}$ 
 & $1.4\pm0.8\times10^{-3}$ & $-20.7\pm0.2$ & $-2.2$(fix) 
 & $2.8\pm1.7\times10^{-4}$
 & $2.7\pm1.8\times10^{25}$ & $2.4\pm1.4\times10^{26}$ \\
$Riz$-LBG & $4.9^{+0.3}_{-0.3}$ 
 & $\simeq 6.4\times10^{-4}$ & $-20.6$(fix) & $-2.2$(fix) 
 & $\simeq 1.1\times10^{-4}$
 & $\simeq 1.1\times10^{25}$ & $\simeq 1.1\times10^{26}$ \\
\tableline
$BRi$-LBG & $4.0^{+0.5}_{-0.5}$ 
 & $2.8\pm0.2\times10^{-3}$ & $-20.6\pm0.1$ & $-1.6$(fix) 
 & $1.8\pm0.1\times10^{-3}$
 & $1.1\pm0.1\times10^{26}$ & $2.4\pm0.2\times10^{26}$ ($4.3\pm0.3\times10^{26}$)\\
$Viz$-LBG & $4.7^{+0.5}_{-0.5}$ 
 & $2.4\pm1.0\times10^{-3}$ & $-20.3\pm0.2$ & $-1.6$(fix) 
 & $2.7\pm1.2\times10^{-4}$
 & $2.7\pm1.3\times10^{25}$ & $1.6\pm0.7\times10^{26}$ ($2.9\pm1.2\times10^{26}$)\\
$Riz$-LBG & $4.9^{+0.3}_{-0.3}$ 
 & $\simeq 1.0\times10^{-3}$ & $-20.3$(fix) & $-1.6$(fix) 
 & $\simeq 1.1\times10^{-4}$
 & $\simeq 1.0\times10^{25}$ & $\simeq 7.0\times10^{25}$ ($\simeq 1.2\times10^{26}$)\\
\tableline
LAE\tablenotemark{\ddag} & $4.86^{+0.03}_{-0.03}$ 
 & $1.9\times10^{-3}$ & $-20.0$ & $-1.6$(fix) 
 & $1.5\times10^{-3}$
 & $5.0\times10^{25}$ & $9.6\times10^{25}$
\enddata
\tablenotetext{\dagger}{
Number density ($n^{\rm obs}$) and 
luminosity density ($\rho_{\rm UV}^{\rm obs}$ ) down
to the observed limiting magnitudes;
$M_{\rm 1700}=-19.8$ ($i'_{\rm ap}=26.5$) for $BRi$-LBGs,
$M_{\rm 1700}=-20.5$ ($z'_{\rm ap}=26.0$) for $Viz$-LBGs,
$M_{\rm 1700}=-20.5$ ($z'_{\rm ap}=26.0$) for $Riz$-LBGs, and
$M_{\rm 1700}=-19.03$ ($i'_{\rm ap}=27.5$) for LAEs,
where $i'_{\rm ap}$ and $z'_{\rm ap}$ are $2''\phi$ aperture 
magnitudes in $i'$ and $z'$, respectively, 
and $M_{\rm 1700}$ is rest-frame 1700\AA\ absolute 
magnitude after aperture correction and k-correction
(see text for details).
}
\tablenotetext{\dagger\dagger}{
Total luminosity density ($\rho_{\rm UV}^{\rm total}$) 
calculated by integrating the luminosity function 
down to $L=0.1L^*$ assuming $\alpha=-2.2$ or $-1.6$.
Values in parentheses are upper limits 
which are calculated by integrating the luminosity function 
down to $L=0$ assuming $\alpha=-1.6$.
}
\tablenotetext{\ddag}{
Lyman $\alpha$ emitters at $z=4.86\pm 0.03$ obtained by
SDS II. We use the best-fit 
UV-lumonisity function given in SDS II
to calculate the UV-luminosity density.
}
\tablecomments{
We refer to the values of
$BRi$-LBGs and $Viz$-LBGs 
as those of $z=4$ and $z=5$ LBGs,
since the values for $Riz$-LBGs are 
quite uncertain. 
In sections \ref{sec:evolution_star}
and \ref{sec:evolution_contribution}, 
we use values obtained for $\alpha=-1.6$.
}


\end{deluxetable}

\clearpage




\begin{thebibliography}{}
\bibitem[Adelberger \& Steidel(2000)]{adelberger2000} Adelberger, 
K.~L.~\& Steidel, C.~C.\ 2000, \apj, 544, 218 
\bibitem[Ascasibar, Yepes, Gottl{\" o}ber, \& M{\" 
u}ller(2002)]{ascasibar2002} Ascasibar, Y., Yepes, G., Gottl{\" 
o}ber, S., \& M{\" u}ller, V.\ 2002, \aap, 387, 396 
\bibitem[Baugh, Cole, Frenk, \& Lacey(1998)]{baugh1998} Baugh, 
C.~M., Cole, S., Frenk, C.~S., \& Lacey, C.~G.\ 1998, \apj, 498, 504 
\bibitem[Ben{\'{\i}}tez(2000)]{benitez2000} Ben{\'{\i}}tez, N.\ 
2000, \apj, 536, 571 
\bibitem[Becker et al.(2001)]{becker2001} Becker, R.~H.~et al.\ 
2001, \aj, 122, 2850 
\bibitem[Bertin \& Arnouts(1996)]{bertin1996} Bertin, E.\ \& Arnouts, S.\ 
1996, \aaps, 117, 393
\bibitem[Bland-Hawthorn \& Maloney(1999)]{bland-hawthorn1999} 
Bland-Hawthorn, J.~\& Maloney, P.~R.\ 1999, \apjl, 510, L33 
\bibitem[Brinchmann \& Ellis(2000)]{brinchmann2000} Brinchmann, J.~\& 
Ellis, R.~S.\ 2000, \apjl, 536, L77 
\bibitem[Bruzual \& Charlot (1993)] {bruzual1993} Bruzual A., G.\ \& 
Charlot, S.\ 1993, \apj, 405, 538
\bibitem[Bruzual \& Charlot (2003)] {bruzual2003} Bruzual A., G.\ \& 
Charlot, S.\ MNRAS, submitted
\bibitem[Calzetti, Kinney, \& Storchi-Bergmann (1994)]{calzetti1994} 
Calzetti, D., Kinney, A. L. \& Storchi-Bergmann, T. 1994, \apj, 429, 582 
\bibitem[Calzetti et al.(2000)]{calzetti2000} Calzetti, D., Armus, 
L., Bohlin, R.~C., Kinney, A.~L., Koornneef, J., \& Storchi-Bergmann, T.\ 
2000, \apj, 533, 682 
\bibitem[Calzetti(2001)]{calzetti2001} Calzetti, D.\ 2001, \pasp, 
113, 1449 
\bibitem[Chen et al.(2003)]{chen2003} Chen, H.~et al.\ 2003, 
\apj, 586, 745 
\bibitem[Cohen(2002)]{cohen2002} Cohen, J.~G.\ 2002, \apj, 567, 672 
\bibitem[Cole et al.(2001)]{cole2001} Cole, S.~et al.\ 2001, 
\mnras, 326, 255 
\bibitem[Coleman, Wu, \& Weedman(1980)]{coleman1980} Coleman, G.~D., Wu, 
C.-C., \& Weedman, D.~W.\ 1980, \apjs, 43, 393
\bibitem[Connolly et al.(1995)]{connolly1995} Connolly, A.~J., Csabai, I., 
Szalay, A.~S., Koo, D.~C., Kron, R.~G., \& Munn, J.~A.\ 1995, \aj, 110, 2655
\bibitem[Cowie, Songaila, Hu, \& Cohen(1996)]{cowie1996} Cowie, 
L.~L., Songaila, A., Hu, E.~M., \& Cohen, J.~G.\ 1996, \aj, 112, 839 
\bibitem[Dickinson(2000)]{dickinson2000} Dickinson, M.\ 2000, Royal 
Society of London Philosophical Transactions Series A, 358, 2001 
(astro-ph/0004028)
\bibitem[Dickinson, Papovich, Ferguson, \& Budav{\' 
a}ri(2003)]{dickinson2003} Dickinson, M., Papovich, C., Ferguson, 
H.~C., \& Budav{\' a}ri, T.\ 2003, \apj, 587, 25 
\bibitem[Fan et al.(2002)]{fan2002} Fan, X., Narayanan, V.~K., 
Strauss, M.~A., White, R.~L., Becker, R.~H., Pentericci, L., \& Rix, H.\ 
2002, \aj, 123, 1247 
\bibitem[Ferguson, Dickinson, \& Papovich(2002)]{ferguson2002} 
Ferguson, H.~C., Dickinson, M., \& Papovich, C.\ 2002, \apjl, 569, L65 
\bibitem[Fern\'andez-Soto, Lanzetta, \& Yahil(1999)]{fernandez-soto1999}
{Fern{\' a}ndez-Soto}, A. and {Lanzetta}, K.\ M. and {Yahil}, A.\ 1999, 
\apj, 513, 34
\bibitem[Fontana et al.(2000)]{fontana2000} Fontana, A., D'Odorico, S., 
Poli, F., Giallongo, E., Arnouts, S., Cristiani, S., Moorwood, A., \& 
Saracco, P.\ 2000, \aj, 120, 2206
\bibitem[Franx et al.(2003)]{franx2003} Franx, M.~et al.\ 2003, 
\apjl, 587, L79 
\bibitem[Fukugita, Shimasaku, \& Ichikawa(1995)]{fukugita1995} 
Fukugita, M., Shimasaku, K., \& Ichikawa, T.\ 1995, \pasp, 107, 945 
\bibitem[Furusawa et al.(2000)]{furusawa2000} Furusawa, H., Shimasaku, K., 
Doi, M., \& Okamura, S.\ 2000, \apj, 534, 624
\bibitem[Furusawa (2002)]{furusawa2002} Furusawa, H.\ 2002, Ph.D.~Thesis, University of Tokyo
\bibitem[Gallego, Zamorano, Aragon-Salamanca, \& 
Rego(1995)]{gallego1995} Gallego, J., Zamorano, J., 
Aragon-Salamanca, A., \& Rego, M.\ 1995, \apjl, 455, L1 
\bibitem[Giallongo, Cristiani, D'Odorico, \& 
Fontana(2002)]{giallongo2002} Giallongo, E., Cristiani, S., 
D'Odorico, S., \& Fontana, A.\ 2002, \apjl, 568, L9 
\bibitem[Glazebrook et al.(1999)]{glazebrook1999} Glazebrook, K., 
Blake, C., Economou, F., Lilly, S., \& Colless, M.\ 1999, \mnras, 306, 843 
\bibitem[Gnedin \& Ostriker(1997)]{gnedin1997} Gnedin, N.~Y.~\& 
Ostriker, J.~P.\ 1997, \apj, 486, 581 
\bibitem[Gunn \& Stryker(1983)]{gunn1983} Gunn, J.~E.~\& Stryker, L.~L.\ 
1983, \apjs, 52, 121
\bibitem[Gwyn \& Hartwick(1996)]{gwyn1996} Gwyn, S.~D.~J.~\& 
Hartwick, F.~D.~A.\ 1996, \apjl, 468, L77 
\bibitem[Hernquist \& Springel(2003)]{hernquist2003} Hernquist, L.~\& 
Springel, V.\ 2003, \mnras, 341, 1253 
\bibitem[Hu et al.(2002)]{hu2002} Hu, E.~M., Cowie, L.~L., 
McMahon, R.~G., Capak, P., Iwamuro, F., Kneib, J.-P., Maihara, T., \& 
Motohara, K.\ 2002, \apjl, 568, L75 
\bibitem[Hurwitz, Jelinsky, \& Dixon(1997)]{hurwitz1997} Hurwitz, 
M., Jelinsky, P., \& Dixon, W.~V.~D.\ 1997, \apjl, 481, L31 
\bibitem[Iwata et al.(2003)]{iwata2003} Iwata, I., Ohta, K., 
Tamura, N., Ando, M., Wada, S., Watanabe, C., Akiyama, M., \& Aoki, K.\ 
2003, \pasj, 55, 415 
\bibitem[Kashikawa et al.(2002)]{kashikawa2002} Kashikawa, N.~et al.\ 
2002, \pasj, 54, 819 
\bibitem[Kashikawa et al.(2003)]{kashikawa2003} Kashikawa, N.~et al.\ 
2003, \aj, 125, 53 (SDS III)
\bibitem[Kauffmann, Colberg, Diaferio, \& White(1999)]{kauffmann1999} 
Kauffmann, G., Colberg, J.~M., Diaferio, A., \& White, S.~D.~M.\ 1999, 
\mnras, 307, 529 
\bibitem[Kodaira et al.(2003)]{kodaira2003} Kodaira, K.~et al.\ 
2003, \pasj, 55, L17 
\bibitem[Labb{\' e} et al.(2003)]{labbe2003} Labb{\' e}, I.~et 
al.\ 2003, \aj, 125, 1107 
\bibitem[Landolt(1992)]{landolt1992} Landolt, A.~U.\ 1992, \aj, 
104, 340 
\bibitem[Lanzetta, Yahil, \& Fernandez-Soto(1996)]{lanzetta1996} 
Lanzetta, K.~M., Yahil, A., \& Fernandez-Soto, A.\ 1996, \nat, 381, 759 
\bibitem[Le Fevre et al.(1996)]{lefevre1996} Le Fevre, O., Hudon, 
D., Lilly, S.~J., Crampton, D., Hammer, F., \& Tresse, L.\ 1996, \apj, 461, 
534 
\bibitem[Leitherer, Ferguson, Heckman, \& 
Lowenthal(1995)]{leitherer1995b} Leitherer, C., Ferguson, H.~C., 
\bibitem[Lowenthal et al.(1997)]{lowenthal1997} Lowenthal, J.~D.~et al.\ 
1997, \apj, 481, 673
\bibitem[Madau(1995)]{madau1995} Madau, P.\ 1995, \apj, 441, 18
\bibitem[Madau et al.(1996)]{madau1996} Madau, P., Ferguson, 
H.~C., Dickinson, M.~E., Giavalisco, M., Steidel, C.~C., \& Fruchter, A.\ 
1996, \mnras, 283, 1388 
\bibitem[Madau, Pozzetti, \& Dickinson(1998)]{madau1998} Madau, P., 
Pozzetti, L., \& Dickinson, M.\ 1998, \apj, 498, 106
\bibitem[Madau, Haardt, \& Rees(1999)]{madau1999} Madau, P., 
Haardt, F., \& Rees, M.~J.\ 1999, \apj, 514, 648 
\bibitem[Maihara et al.(2001)]{maihara2001} Maihara, T.~et al.\ 2001, 
\pasj, 53, 25 (SDS I)
\bibitem[Massarotti, Iovino, Buzzoni, \& 
Valls-Gabaud(2001)]{massarotti2001} Massarotti, M., Iovino, A., 
Buzzoni, A., \& Valls-Gabaud, D.\ 2001, \aap, 380, 425 
\bibitem[Meurer, Heckman, \& Calzetti(1999)]{meurer1999} Meurer, 
G.~R., Heckman, T.~M., \& Calzetti, D.\ 1999, \apj, 521, 64 
\bibitem[Miralda-Escude \& Rees(1998)]{miralda-escude1998} Miralda-Escude, 
J.~\& Rees, M.~J.\ 1998, \apj, 497, 21 
\bibitem[Miyazaki et al.(2002)]{miyazaki2002} Miyazaki, S.~et al.\ 
2002, \pasj, 54, 833 
\bibitem[Monet et al.(1998)]{monet1998} Monet, D.\ E.\ A.\ 1998, 
The PMM USNO-A2.0 Catalog.
\bibitem[Motohara et al.(2002)]{motohara2002} Motohara, K.~et al.\ 
2002, \pasj, 54, 315 
\bibitem[Nagamine, Cen, \& Ostriker(2000)]{nagamine2000} Nagamine, 
K., Cen, R., \& Ostriker, J.~P.\ 2000, \apj, 541, 25 
\bibitem[Oke(1974)]{oke1974} Oke, J.~B.\ 1974, \apjs, 27, 21 
\bibitem[Oke(1990)]{oke1990} Oke, J.~B.\ 1990, \aj, 99, 1621 
\bibitem[Ouchi, Yamada, Kawai, \& Ohta(1999)]{ouchi1999} Ouchi, 
M., Yamada, T., Kawai, H., \& Ohta, K.\ 1999, \apjl, 517, L19 
\bibitem[Ouchi et al.(2001)]{ouchi2001a} Ouchi, M.~et al.\ 2001, \apjl, 558, L83
\bibitem[Ouchi et al.(2003a)]{ouchi2003a} Ouchi, M.~et al.\ 2003a, 
\apj, 582, 60 (SDS II)
\bibitem[Ouchi et al.(2003b)]{ouchi2003c}  Ouchi, M.\ 2003b, Ph.D.~Thesis, University of Tokyo 
\bibitem[Ouchi et al.(2003c)]{ouchi2003d}  Ouchi, M.\ 2003c, submitted to ApJ (astro-ph/0309657) (SDS VI)
\bibitem[Papovich, Dickinson, \& Ferguson(2001)]{papovich2001} 
Papovich, C., Dickinson, M., \& Ferguson, H.~C.\ 2001, \apj, 559, 620 
\bibitem[Pascarelle, Windhorst, \& Keel(1998)]{pascarelle1998} Pascarelle, 
S.~M., Windhorst, R.~A., \& Keel, W.~C.\ 1998, \aj, 116, 2659
\bibitem[Pettini et al. (1998)]{pettini1998} Pettini, M., Kellogg, 
M., Steidel, C. C., Dickinson, M., Adelberger, K. L. \& Giavalisco, M. 
1998, \apj, 508, 539
\bibitem[Pettini et al.(2001)]{pettini2001} Pettini, M., Shapley, 
A.~E., Steidel, C.~C., Cuby, J., Dickinson, M., Moorwood, A.~F.~M., 
Adelberger, K.~L., \& Giavalisco, M.\ 2001, \apj, 554, 981 
\bibitem[Pettini et al.(2002)]{pettini2002} Pettini, M., Rix, 
S.~A., Steidel, C.~C., Adelberger, K.~L., Hunt, M.~P., \& Shapley, A.~E.\ 
2002, \apj, 569, 742 
\bibitem[Poli et al.(2001)]{poli2001} Poli, F., Menci, N., 
Giallongo, E., Fontana, A., Cristiani, S., \& D'Odorico, S.\ 2001, \apjl, 
551, L45 
\bibitem[Rudnick et al.(2001)]{rudnick2001} Rudnick, G.~et al.\ 
2001, \aj, 122, 2205 
\bibitem[Sawicki, Lin, \& Yee(1997)]{sawicki1997} Sawicki, M.~J., Lin, H., 
\& Yee, H.~K.~C.\ 1997, \aj, 113, 1
\bibitem[Sawicki \& Yee(1998)]{sawicki1998} Sawick, M., \& Yee,
H. K. C. 1998, AJ, 115, 1329
\bibitem[Schechter(1976)]{schechter1976} Schechter, P.\ 1976, \apj, 
203, 297 
\bibitem[Schlegel, Finkbeiner, \& Davis(1998)]{schlegel1998} 
Schlegel, D.~J., Finkbeiner, D.~P., \& Davis, M.\ 1998, \apj, 500, 525 
\bibitem[Shapley et al.(2001)]{shapley2001} Shapley, A.~E., Steidel, C.~C., 
Adelberger, K.~L., Dickinson, M., Giavalisco, M., \& Pettini, M.\ 2001, 
\apj, 562, 95
\bibitem[Shimasaku et al.(2003)]{shimasaku2003} Shimasaku, K.~et al.\ 
2003, \apjl, 586, L111 (SDS IV)
\bibitem[Spergel et al.(2003)]{spergel2003} Spergel, D.~N.~et al.\ 
2003, ApJ in press, (astro-ph/0302209)
\bibitem[Steidel et al.(1996a)]{steidel1996a} Steidel, C.\ C., Giavalisco, 
M., Pettini, M., Dickinson, M., \& Adelberger, K.\ L.\ 1996a, \apjl, 462, L17
\bibitem[Steidel et al.(1996b)]{steidel1996b} Steidel, C.~C., Giavalisco, 
M., Dickinson, M., \& Adelberger, K.~L.\ 1996b, \aj, 112, 352
\bibitem[Steidel et al.(1999)]{steidel1999} Steidel, C.\ C., Adelberger, 
K.\ L., Giavalisco, M., Dickinson, M., \& Pettini, M.\ 1999, \apj, 519, 1
\bibitem[Steidel, Pettini, \& Adelberger(2001)]{steidel2001} 
Steidel, C.~C., Pettini, M., \& Adelberger, K.~L.\ 2001, \apj, 546, 665 
\bibitem[Steidel et al.(2002)]{steidel2002} Steidel, C.~C., Hunt, 
M.~P., Shapley, A.~E., Adelberger, K.~L., Pettini, M., Dickinson, M., \& 
Giavalisco, M.\ 2002, \apj, 576, 653 
\bibitem[Steidel et al.(2003)]{steidel2003} Steidel, C.~C., 
Adelberger, K.~L., Shapley, A.~E., Pettini, M., Dickinson, M., \& 
Giavalisco, M.\ 2003, ArXiv Astrophysics e-prints, 5378 (astro-ph/0305378)
\bibitem[Sullivan et al.(2000)]{sullivan2000} Sullivan, M., Treyer, 
M.~A., Ellis, R.~S., Bridges, T.~J., Milliard, B., \& Donas, J.~; 2000, 
\mnras, 312, 442 
\bibitem[Tresse \& Maddox(1998)]{tresse1998} Tresse, L.~\& Maddox, 
S.~J.\ 1998, \apj, 495, 691 
\bibitem[Tumlinson, Giroux, Shull, \& Stocke(1999)]{tumlinson1999} 
Tumlinson, J., Giroux, M.~L., Shull, J.~M., \& Stocke, J.~T.\ 1999, \aj, 
118, 2148 
\bibitem[Vijh, Witt, \& Gordon(2003)]{vijh2003} Vijh, U.~P., 
Witt, A.~N., \& Gordon, K.~D.\ 2003, \apj, 587, 533 
\bibitem[Wang, Bahcall, \& Turner(1998)]{wang1998} Wang, Y., 
Bahcall, N., \& Turner, E.~L.\ 1998, \aj, 116, 2081 
\bibitem[Weinberg, Hernquist, \& Katz(2002)]{weinberg2002} Weinberg, 
D.~H., Hernquist, L., \& Katz, N.\ 2002, \apj, 571, 15 
\bibitem[Yagi et al.(2002)]{yagi2002} Yagi, M., Kashikawa, N., 
Sekiguchi, M., Doi, M., Yasuda, N., Shimasaku, K., \& Okamura, S.\ 2002, 
\aj, 123, 66 
\bibitem[Yahata et al.(2000)]{yahata2000} Yahata, N., Lanzetta, 
K.~M., Chen, H., Fern{\' a}ndez-Soto, A., Pascarelle, S.~M., Yahil, A., \& 
Puetter, R.~C.\ 2000, \apj, 538, 493 
\bibitem[Yan et al.(1999)]{yan1999} Yan, L., McCarthy, P.~J., 
Freudling, W., Teplitz, H.~I., Malumuth, E.~M., Weymann, R.~J., \& Malkan, 
M.~A.\ 1999, \apjl, 519, L47 
\bibitem[Yan, Windhorst, \& Cohen(2003)]{yan2003} Yan, H., 
Windhorst, R.~A., \& Cohen, S.~H.\ 2003, \apjl, 585, L93 

\end{thebibliography}
\end{document}